\documentclass[aps,prd,reprint,superscriptaddress]{revtex4-2}
\pdfoutput=1 

\usepackage{graphicx}
\usepackage{amsmath}
\usepackage{hyperref }
\usepackage{lineno}
\usepackage{gensymb}
\usepackage{makecell}
\usepackage{hhline}
\usepackage{multirow}
\usepackage{subcaption}
\usepackage{ragged2e}
\DeclareCaptionJustification{justified}{\justifying}
\graphicspath{ {images/} }

\begin{document}

\title{Measurement of the Neutron Cross Section on Argon Between 95 and 720 MeV}

\author{S.~Martynenko} 
\email[]{smartynen@bnl.gov}
\affiliation{Brookhaven National Laboratory, Upton, NY 11973, USA}
\affiliation{Department of Physics and Astronomy, Stony Brook University, Stony Brook, NY 11794, USA}
\author{B.~Bhandari} 
\affiliation{Department of Physics, University of Houston, Houston, TX 77204, USA}
\author{J.~Bian} 
\affiliation{Department of Physics and Astronomy, University of California, Irvine, CA 92697, USA}
\author{K.~Bilton} 
\affiliation{Department of Physics, University of California, Davis, CA 95616 }
\author{C.~Callahan} 
\affiliation{Department of Physics and Astronomy, University of Pennsylvania, Philadelphia, PA 19104, USA}
\author{J.~Chaves}
\affiliation{Department of Physics and Astronomy, University of Pennsylvania, Philadelphia, PA 19104, USA}
\author{H.~Chen}
\affiliation{Brookhaven National Laboratory, Upton, NY 11973, USA}
\author{D.~Cline}
\thanks{Deceased}
\affiliation{Department of Physics and Astronomy, University of California, Los Angeles, CA 90095, USA}
\author{R.~L.~Cooper}
\affiliation{Department of Physics, New Mexico State University, Las Cruces, NM 88003, USA}
\author{D.~L.~Danielson} 
\affiliation{Enrico Fermi Institute and Department of Physics, The University of Chicago, Chicago, IL 60637, USA}
\author{J.~Danielson} 
\affiliation{Los Alamos National Laboratory, Los Alamos, NM 87545, USA}
\author{N.~Dokania} 
\affiliation{Department of Physics and Astronomy, Stony Brook University, Stony Brook, NY 11794, USA}
\author{S.~Elliott} 
\affiliation{Los Alamos National Laboratory, Los Alamos, NM 87545, USA}
\author{S.~Fernandes}
\affiliation{Department of Physics and Astronomy, University of Alabama, Tuscaloosa, AL 35487, USA}
\author{S.~Gardiner}
\affiliation{Department of Physics, University of California, Davis, CA 95616 }
\author{G.~Garvey} 
\affiliation{Los Alamos National Laboratory, Los Alamos, NM 87545, USA}
\author{V.~Gehman} 
\affiliation{Lawrence Berkeley National Laboratory, Berkeley, CA 94720, USA}
\author{F.~Giuliani} 
\affiliation{Department of Physics and Astronomy, University of New Mexico, Albuquerque, NM 87131, USA}
\author{S.~Glavin} 
\affiliation{Department of Physics and Astronomy, University of Pennsylvania, Philadelphia, PA 19104, USA}
\author{M.~Gold} 
\affiliation{Department of Physics and Astronomy, University of New Mexico, Albuquerque, NM 87131, USA}
\author{C.~Grant}
\affiliation{Department of Physics, Boston University, Boston, MA 02215, USA}
\author{E.~Guardincerri} 
\affiliation{Los Alamos National Laboratory, Los Alamos, NM 87545, USA}
\author{T.~Haines} 
\affiliation{Los Alamos National Laboratory, Los Alamos, NM 87545, USA}
\author{A.~Higuera} 
\affiliation{Department of Physics and Astronomy, Rice University, Houston, TX 77005, USA}
\author{J.~Y.~Ji} 
\affiliation{Department of Physics and Astronomy, Stony Brook University, Stony Brook, NY 11794, USA}\author{R.~Kadel}
\affiliation{Lawrence Berkeley National Laboratory, Berkeley, CA 94720, USA}
\author{N.~Kamp} 
\affiliation{Los Alamos National Laboratory, Los Alamos, NM 87545, USA}
\author{A.~Karlin} 
\affiliation{Department of Physics and Astronomy, University of Pennsylvania, Philadelphia, PA 19104, USA}
\author{W.~Ketchum} 
\affiliation{Los Alamos National Laboratory, Los Alamos, NM 87545, USA}
\author{L.~W.~Koerner} 
\affiliation{Department of Physics, University of Houston, Houston, TX 77204, USA}
\author{D.~Lee} 
\affiliation{Los Alamos National Laboratory, Los Alamos, NM 87545, USA}
\author{K.~Lee} 
\affiliation{Department of Physics and Astronomy, University of California, Los Angeles, CA 90095, USA}
\author{Q.~Liu} 
\affiliation{Los Alamos National Laboratory, Los Alamos, NM 87545, USA}
\author{S.~Locke} 
\affiliation{Department of Physics and Astronomy, University of California, Irvine, CA 92697, USA}
\author{W.~C.~Louis} 
\affiliation{Los Alamos National Laboratory, Los Alamos, NM 87545, USA}
\author{J.~Maricic} 
\affiliation{Department of Physics and Astronomy, University of Hawaii at Manoa, Honolulu, HI 96822, USA}
\author{E.~Martin} 
\affiliation{Department of Physics and Astronomy, University of California, Los Angeles, CA 90095, USA}
\author{M.~J.~Martinez}
\affiliation{Los Alamos National Laboratory, Los Alamos, NM 87545, USA}

\author{C.~Mauger}
\affiliation{Department of Physics and Astronomy, University of Pennsylvania, Philadelphia, PA 19104, USA}
\author{C.~McGrew} 
\affiliation{Department of Physics and Astronomy, Stony Brook University, Stony Brook, NY 11794, USA}
\author{J.~Medina} 
\affiliation{Los Alamos National Laboratory, Los Alamos, NM 87545, USA}
\author{P.~J.~Medina}
\thanks{Deceased}
\affiliation{Los Alamos National Laboratory, Los Alamos, NM 87545, USA}
\author{A.~Mills} 
\affiliation{Department of Physics and Astronomy, University of New Mexico, Albuquerque, NM 87131, USA}
\author{G.~Mills}
\thanks{Deceased}
\affiliation{Los Alamos National Laboratory, Los Alamos, NM 87545, USA}
\author{J.~Mirabal-Martinez} 
\affiliation{Los Alamos National Laboratory, Los Alamos, NM 87545, USA}
\author{A.~Olivier}
\affiliation{Department of Physics and Astronomy, Louisiana State University, Baton Rouge, LA 70803, USA}
\author{E.~Pantic} 
\affiliation{Department of Physics, University of California, Davis, CA 95616 }
\author{B.~Philipbar}
\affiliation{Department of Physics and Astronomy, University of New Mexico, Albuquerque, NM 87131, USA}
\author{C.~Pitcher} 
\affiliation{Department of Physics and Astronomy, University of California, Irvine, CA 92697, USA}
\author{V.~Radeka} 
\affiliation{Brookhaven National Laboratory, Upton, NY 11973, USA}
\author{J.~Ramsey} 
\affiliation{Los Alamos National Laboratory, Los Alamos, NM 87545, USA}
\author{K.~Rielage} 
\affiliation{Los Alamos National Laboratory, Los Alamos, NM 87545, USA}
\author{M.~Rosen} 
\affiliation{Department of Physics and Astronomy, University of Hawaii at Manoa, Honolulu, HI 96822, USA}
\author{A.~R.~Sanchez}
\affiliation{Los Alamos National Laboratory, Los Alamos, NM 87545, USA}
\author{J.~Shin} 
\affiliation{Department of Physics and Astronomy, University of California, Los Angeles, CA 90095, USA}
\author{G.~Sinnis} 
\affiliation{Los Alamos National Laboratory, Los Alamos, NM 87545, USA}
\author{M.~Smy}
\affiliation{Department of Physics and Astronomy, University of California, Irvine, CA 92697, USA}
\author{W.~Sondheim} 
\affiliation{Los Alamos National Laboratory, Los Alamos, NM 87545, USA}
\author{I.~Stancu}
\affiliation{Department of Physics and Astronomy, University of Alabama, Tuscaloosa, AL 35487, USA}
\author{C.~Sterbenz} 
\affiliation{Los Alamos National Laboratory, Los Alamos, NM 87545, USA}
\author{Y.~Sun}
\affiliation{Department of Physics and Astronomy, University of Hawaii at Manoa, Honolulu, HI 96822, USA}
\author{R.~Svoboda} 
\affiliation{Department of Physics, University of California, Davis, CA 95616 }
\author{C.~Taylor} 
\affiliation{Los Alamos National Laboratory, Los Alamos, NM 87545, USA}
\author{A.~Teymourian} 
\affiliation{Department of Physics and Astronomy, University of California, Los Angeles, CA 90095, USA}
\author{C.~Thorn}
\affiliation{Brookhaven National Laboratory, Upton, NY 11973, USA}
\author{C.~E.~Tull}
\affiliation{Lawrence Berkeley National Laboratory, Berkeley, CA 94720, USA}
\author{M.~Tzanov}
\affiliation{Department of Physics and Astronomy, Louisiana State University, Baton Rouge, LA 70803, USA}
\author{R.~G.~Van de Water} 
\affiliation{Los Alamos National Laboratory, Los Alamos, NM 87545, USA}
\author{D.~Walker}
\affiliation{Department of Physics and Astronomy, Louisiana State University, Baton Rouge, LA 70803, USA}
\author{N.~Walsh}
\affiliation{Department of Physics, University of California, Davis, CA 95616 }
\author{H.~Wang}
\affiliation{Department of Physics and Astronomy, University of California, Los Angeles, CA 90095, USA}
\author{Y.~Wang} 
\affiliation{Department of Physics and Astronomy, University of California, Los Angeles, CA 90095, USA}
\author{C.~Yanagisawa}
\affiliation{Department of Physics and Astronomy, Stony Brook University, Stony Brook, NY 11794, USA}
\author{A.~Yarritu}
\affiliation{Los Alamos National Laboratory, Los Alamos, NM 87545, USA}
\author{J.~Yoo}
\affiliation{University of Illinois Chicago, Chicago, IL 60605, USA}


\collaboration{CAPTAIN Collaboration}
\noaffiliation

\date{\today}

\begin{abstract}
We report an extended measurement of the neutron cross section on argon in the energy range of 95-720 MeV. The measurement was obtained with a 4.3-hour exposure of the Mini-CAPTAIN detector to the WNR/LANSCE beam at LANL. Compared to an earlier analysis of the same data, this extended analysis includes a reassessment of systematic uncertainties, in particular related to unused wires in the upstream part of the detector. Using this information we doubled the fiducial volume in the experiment and increased the statistics by a factor of 2.4. We also shifted the analysis from energy bins to time-of-flight bins. This change reduced the overall considered energy range, but improved the understanding of the energy spectrum of incoming neutrons in each bin. Overall, the new measurements are extracted from a fit to the attenuation of the neutron flux in five time-of-flight regions: 140~ns~-~180~ns, 120~ns~-~140~ns, 112~ns~-~120~ns, 104~ns~-~112~ns, 96~ns~-~104~ns. The final cross sections are given for the flux-averaged energy in each time-of-flight bin: $\sigma(146~\rm{MeV})=0.60^{+0.14}_{-0.14}\pm0.08$(syst) b, $\sigma(236~\rm{MeV})=0.72^{+0.10}_{-0.10}\pm0.04$(syst) b, $\sigma(319~\rm{MeV})=0.80^{+0.13}_{-0.12}\pm0.040$(syst) b, $\sigma(404~\rm{MeV})=0.74^{+0.14}_{-0.09}\pm0.04$(syst) b, $\sigma(543~\rm{MeV})=0.74^{+0.09}_{-0.09}\pm0.04$(syst) b.
\end{abstract}


\maketitle

\section{Introduction}

The Liquid Argon Time Projection Chamber (LArTPC) technology, originally proposed for neutrino detectors \cite{original_tpc}, is used in multiple neutrino experiments \cite{DUNE,SBND,microboone,lariat}. This detection method has high precision and low energy threshold, which together allows highly detailed reconstruction of neutrino events. As a charged particle passes through a medium, it creates ionization. In a LArTPC, an electric field causes the produced electrons to drift to read-out planes. Often these consist of parallel sense wires. The drift time and the position of the hit wires are combined to provide a 3D reconstruction of the event. 

Neutrino interactions produce neutrons in addition to charged particles.  Like neutrinos, neutrons have no electric charge and can't be directly detected. They also carry a considerable amount of energy\cite{Friedland:2018vry,PhysRevD.92.091301}. This energy is missing in the calorimetric measurement adding significant uncertainty to neutrino energy reconstruction and, as a result, neutrino oscillation measurements. Models used to estimate missing energy, including neutrons, have large unconstrained uncertainties. In order to improve neutrino energy reconstruction in LArTPCs, precise measurements of the neutron cross section in liquid argon are needed for a broad range of energies.  Prior to the effort of the CAPTAIN (The Cryogenic Apparatus for Precision Tests of Argon Interactions with Neutrinos) collaboration, neutron-argon cross section data was only published up to 50 MeV of kinetic energy\cite{old_neutron}. CAPTAIN reported its first measurement of the neutron cross section between 100 and 800 MeV in 2017\cite{mypaper}. In this work, we significantly extend the earlier measurement. We carefully studied multiple systematic uncertainties and their effect on the measurement, especially at the upstream part of the detector. We improved the statistics for the measurement by a factor of 2.4 by doubling the fiducial volume.  Moreover, we switched to time-of-flight bins in the analysis for better understanding of the incoming neutron energies for cross section calculation.  
 
Since we extracted the measurement from a fit to the attenuation of the neutron flux, it should be perceived as a beam depletion cross section. In other words, it is the cross section to remove a neutron from a 50~mm radius circular area surrounding the best fit  beam center. 

This paper is organized as follows. First, Section~\ref{sec:ExpSetuup} describes
the key aspects of the neutron beam and the detector used for this measurement. Second, Section~\ref{sec:Rec} describes the event reconstruction used in the experiment. Section~\ref{sec:Perf} reports the study of the  detector performance. Next, Section~\ref{sec:Sim} introduces the Monte Carlo (MC) simulation process. Finally, Section~\ref{sec:Data} focuses on the event selection and data analysis strategy. The results are presented in Section~\ref{sec:Data} and followed by conclusions in Section~\ref{sec:Summ}.

\section{Experiment Setup}
\label{sec:ExpSetuup}

This section includes descriptions of the Mini-CAPTAIN detector, the WNR facility at Los Alamos National Laboratory, and the neutron beam provided by LANSCE\cite{lanlfac}. The section is finished with a discussion of the data set, collected during the Mini-CAPTAIN neutron run during the summer of 2017. A more detailed description can be found in~\cite{taylor2020mini}. 

\subsection{Mini-CAPTAIN detector}

\begin{figure}[h]
  \centering
  \includegraphics[width=0.8\linewidth]{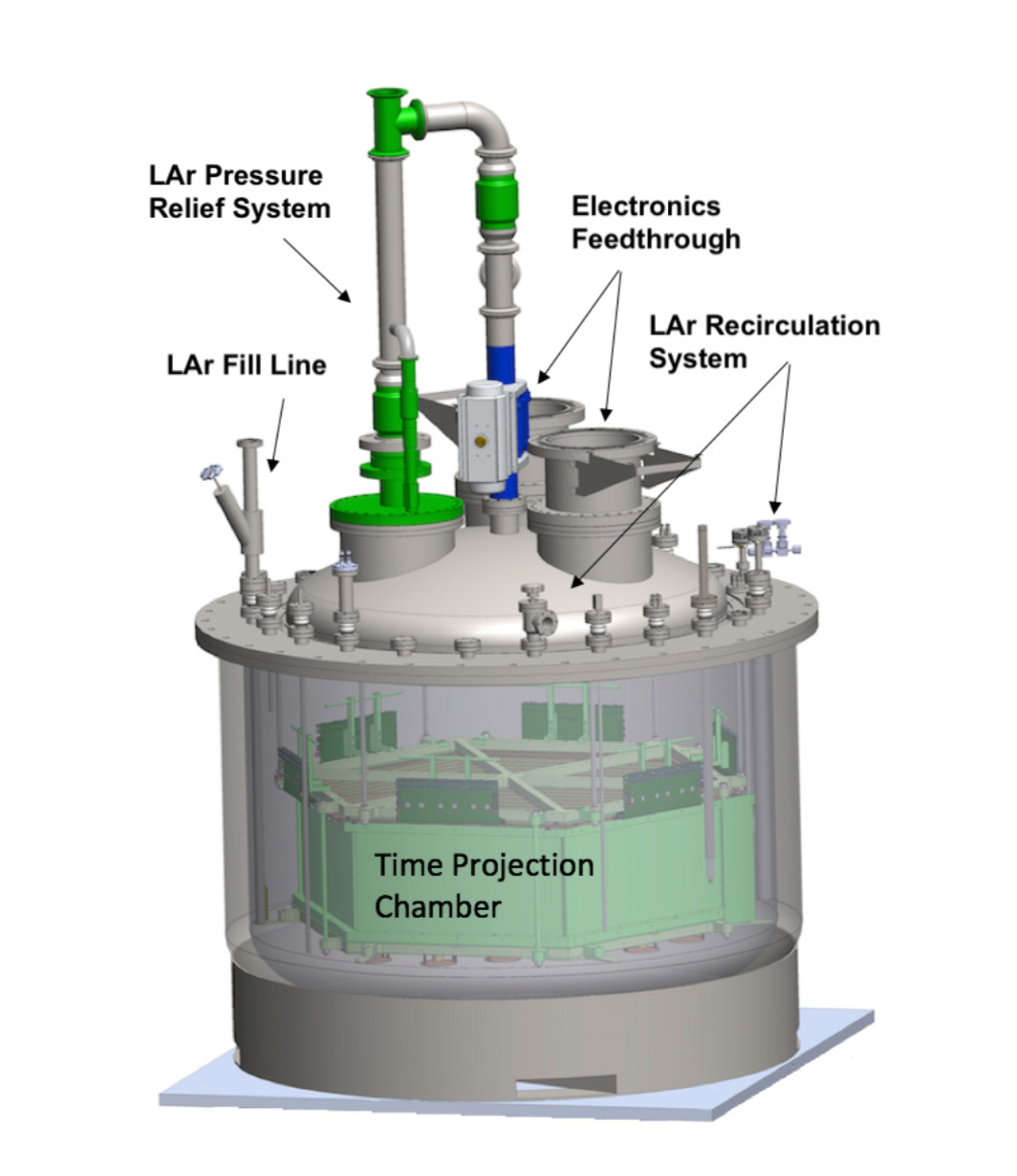}  

    \caption{The schematic of the Mini-CAPTAIN detector\cite{taylor2020mini}.}
    \label{fig:DETECTOR}
\end{figure}

The Mini-CAPTAIN is a hexagonal LArTPC with a photon detection system (PDS). The latter is used to measure the times of neutron interactions. We combine it with an initial time for each neutron derived from the radio-frequency (RF) signal picked up by a copper coil at the beam target location to measure the neutron time-of-flight. 

\subsection{TPC design}

\begin{figure}
\begin{center}
\includegraphics[width=0.42\textwidth]{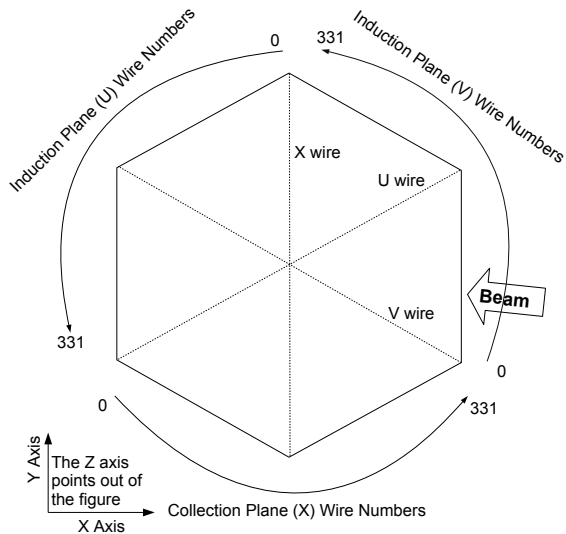}
\end{center}
\caption{The schematic of the positions of the wires with respect to the beam in Mini-CAPTAIN. The hexagonal plane of the TPC lies in the XY plane with the zero coordinate in the middle of the cryostat. The Z-axis pierces the detector vertically with zero at the top of the cryostat.}
\label{fig:TPCcoord}
\end{figure}

The schematic drawing of the detector is shown in Fig.~\ref{fig:DETECTOR}. The TPC has a hexagonal shape with an apothem of 50~cm and 32~cm of vertical drift between the cathode at the bottom and the anode at the top. The charged particles traveling through liquid argon create ionization electrons. The 500~V/cm electric field is applied across the TPC volume for the ionization electrons to drift toward wire planes (X, U, and V). The provided electric field results in 1.6~mm$/\mu$s electron drift velocity. The so-called X wire plane is the collection wire plane with wires positioned almost perpendicular to the neutron beam (or X-axis in the coordinate system used in the analysis). The other two planes are induction wire planes called U and V. Wires on these planes are positioned $\pm60^\circ$ with respect to the collection wire plane or $\pm30^\circ$ with respect to the X-axis. Each wire plane has 337 copper-beryllium wires 75~$\mu$m in diameter. The distance between wires is 3.125~mm across all wire planes. The coordinate system used in the analysis and the schematic of the positions of the wires with respect to the beam are shown in Fig.~\ref{fig:TPCcoord}. The center of the XY plane is aligned with the center of the cryostat. The zero of the Z-axis is located at the top of the cryostat. 

Liquid argon purification is performed in three stages: inline filter, gas recirculation system, and a liquid argon purification system from Criotec Impianti\cite{criotech}. The combination of all three purification techniques provided 0.3 and 1.5 ppb concentrations of H$_2$O and O$_2$, respectively, and a sufficient electron lifetime for the measurement. A detailed discussion of the electron lifetime is presented in Section~\ref{subs:Elife}.

\subsection{PDS design}

The PDS measures the light from neutron interactions to establish the event time and thereby neutron energy. The LArTPC readout time is O(100~$\mu$s) because of the slow drift velocity while the scintillation light detection time resolution is a few ns.

The photon detection system consists of 24 Hamamatsu R8520-506 MOD photomultiplier tubes (PMTs), each approximately 1” $\times$ 1” $\times$ 1” in size. All PMTs have a borosilicate glass window and a special bialkali photocathode capable of operation at liquid argon temperatures (87~K). The PMTs are mounted on both the top and bottom of the cryostat. The signal from the PMTs is digitized by three CAEN V1720 digitizers with 4~ns resolution. Each digitizer is taking data from seven PMTs as well as the RF signal.

\subsection{Neutron beam}

The Mini-CAPTAIN detector was deployed at the Los Alamos Neutron Science Center (LANSCE) in the Target 4, flight path 15R (4FP15R) beamline of the Weapon Neutron Research (WNR) facility during the summer of 2017. The Target 4 facility uses a proton beam and tungsten target to produce neutrons\cite{nowicki2017alamos}. The neutron energy spectrum covers energies from 1~MeV up to 800~MeV.

Two beam structures were provided by LANSCE. For the regular beam operation, the beam came in macropulses that were 625~$\mu$s wide, separated by a minimum of 8.3~ms. Each macropulse consisted of micropulses, which were 100~ps wide and separated by 1.8~$\mu$s. The second beam mode was provided specifically for the CAPTAIN experiment. The overall macropulse structure of the beam stayed the same. However, the number of micropulses inside each macropulse was reduced to three. Thus, micropulses were separated by 199~$\mu$s, but had the same 100~ps width as in the regular beam mode.

Aside from the reduced amount of micropulses per macropulse, additional reduction of the neutron flux was required to prevent event pile-up issues in the TPC where the drift time was 200~$\mu$s. We operated shutters mounted on the beamline to reach a neutron flux of about one neutron per macropulse. 

\subsection{Detector triggering and analysis data set}

We collect data separately for the TPC and PDS systems. We synchronize the two data streams during the event reconstruction stage as described in Section~\ref{sec:Rec}. Both systems trigger by the beam RF pulse. The RF pulse is a signal from the copper coil around the target location indicating the time protons strike the target. The TPC data acquisition window of 4.75~ms is designed to include the 625~$\mu$s macropulse along with 1.85~ms prior to and 2.3~ms after the trigger time. The PDS data acquisition window is set to 8~$\mu$s and triggered with the arrival of the RF pulse as well. However, the PDS could potentially trigger independently from the RF pulse if enough light is seen in the detector. 

The data set used for the analysis was obtained during the CAPTAIN-specific beam structure on August 31, 2017. We perform the analysis with approximately 4.3~hours of neutron data. Due to the specific trigger setup, based on the combination of the RF pulse and light from the neutron interactions, the background from cosmics is negligible. However, the cosmic data was collected separately during these hours and is used to study detector performance.

\section{Reconstruction}
\label{sec:Rec}

We use the calibrated ionization signal from the wire planes to reconstruct the three-dimensional tracks of charged particles through the TPC. The thermal noise is removed from the ionization signal in each wire using a Wiener filter. It is used assuming the ionization signal is smooth as a function of frequency. The peaks that remain after the filtering are used to make hits for track reconstruction. In particular, each hit has an associated time, charge integral, and location (the spatial coordinates of the wire). 

Neutron interactions in liquid argon produce mostly protons, pions, muons, and electrons, all of which leave linear tracks in the detector.  Consequently, we designed the reconstruction algorithm to handle straight objects with possible small bands from secondary interactions. In each plane, hits located along straight lines are grouped together using proximity clustering \cite{patent:3069654,dbscan}. Additional hits were added to these track seeds if they were in a 40~mm box around the end of the track. This additional step improved the efficiency to reconstruct tracks that have gone through multiple scattering. Track candidates are then associated with candidates in other planes. A 3D track is formed from a track candidate in the collection plane and at least one track candidate from either induction plane. Finally, we use an SIR (sequential importance resampling) particle filter with forward/backward filtering on track candidates to obtain the beginning and end points of the track.

The reconstruction for the PDS is performed for each event (8.4~$\mu$s window with 4~ns sampling). First, the mean value for all samples in each PMT is calculated. This serves as a baseline for peak finding. Next, the average noise for the PMT is measured looking at fluctuation between two consecutive samples. The standard deviation of the distribution defines the noise RMS for the specific PMT. We define a peak candidate as a sample with a value of more than 1.5 standard deviations above the mean PMT sample value. The charge in photoelectrons ($C_{PE}$) for each peak is calculated using:

\begin{equation}
    C_{PE}=\frac{S-M}{G},
\end{equation}
where $S$ is the sample value of the peak, $M$ is the mean sample value for the PMT, and $G$ is the calibration gain constant for the PMT \cite{taylor2020mini}. Hits are constructed based on the found peaks. We set the threshold to define a hit at 0.4~photoelectrons, which eliminates noise while selecting a single photoelectron signal. 

All found hits in the event are stored with the reconstructed time assigned to them. If the RF pulse is present in this event, the reconstructed time is set to be the difference between the hit's peak time and the RF pulse starting time for the corresponding digitizer. If there is no RF pulse, the time is set to be just the time of the peak of each hit with respect to self-trigger time. We use reconstructed time to separate hits into 16~ns blocks (4 samples). If two or more PMTs observe a signal within 16~ns of each other, we call it a coincidence. The time interval between such coincidences and the RF signal time define the neutron energy spectrum, as the coincidence time measures the neutron's interaction time in the detector. For cosmic particle interactions, there is no RF signal, and the coincidence time measures the particle interaction time.

Finally, each TPC event is associated with one or several corresponding PDS events. Since the beam micropulses occur every 199~$\mu$s, all PDS events occurring within $\pm100~\mu$s of a TPC event are associated with that TPC event.

\section{Detector Performance}
\label{sec:Perf}

Detector performance is studied using a sample of long cosmic muon tracks defined as a straight line traversing through the entire drift distance (32~cm). We also require the muon track to be outside of the beam time window. 

\subsection{Electron Lifetime}
\label{subs:Elife}

The experiment is based around ionization electrons drifting through liquid argon to the wire planes. Thus, it is essential for the argon to be purified against electro-negative impurities. We show the result of the purification process by studying electron lifetime in the detector.

The analysis is done for the three wire planes separately. The deposited charge is calculated as a sum of charges of unique hits in a given time frame. Charges are corrected based on the angle between the track and the wires.

\begin{figure}[h]
    \begin{subfigure}[b]{0.462\textwidth}
        \centering
        \includegraphics[width=1.0\textwidth]{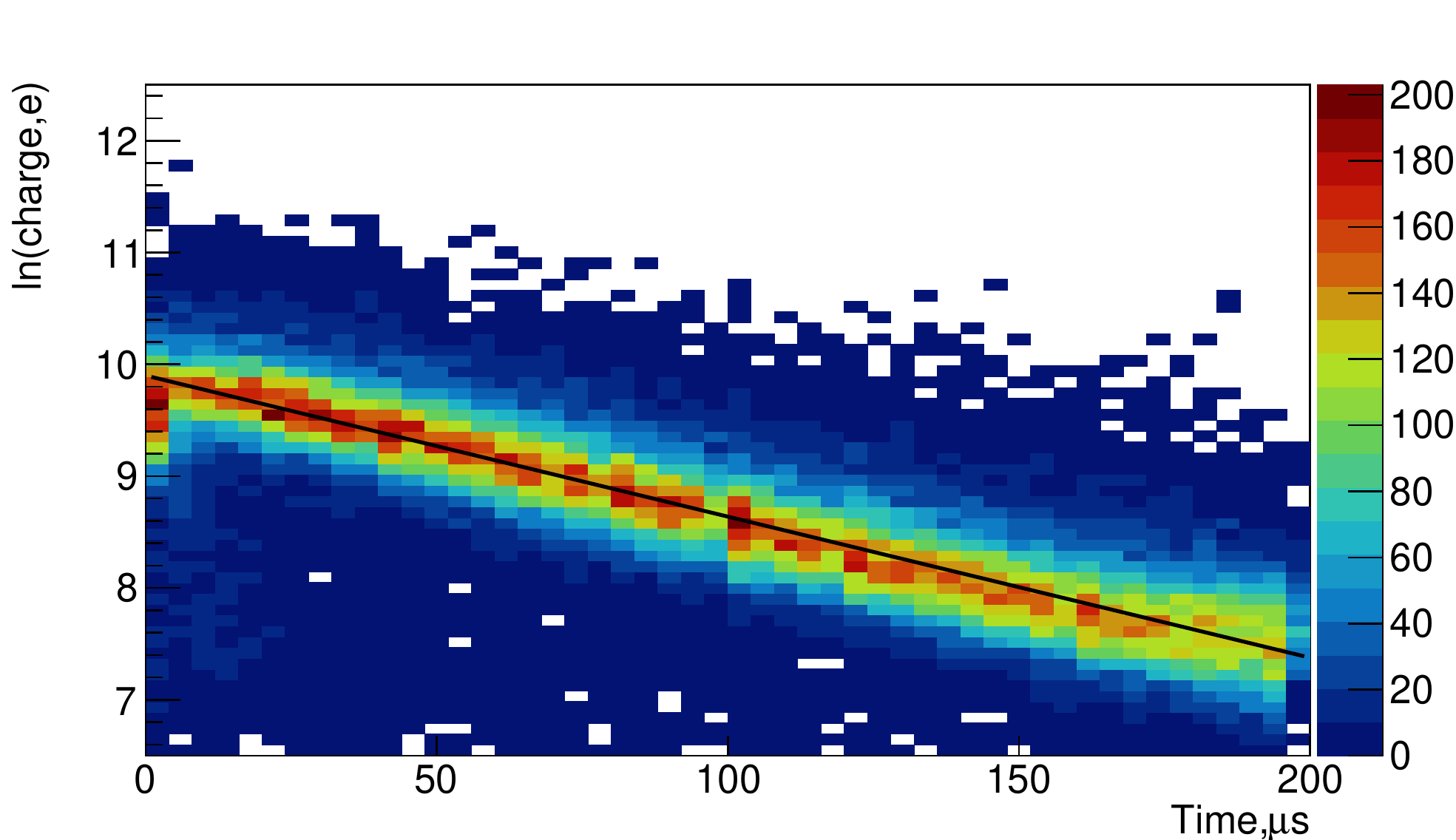}
        \caption{\centering X plane hit charge vs time}
    \label{fig:EliftimeX}
    \end{subfigure}\\

    \begin{subfigure}[b]{0.48\textwidth}
        \centering
        \includegraphics[width=0.96\textwidth]{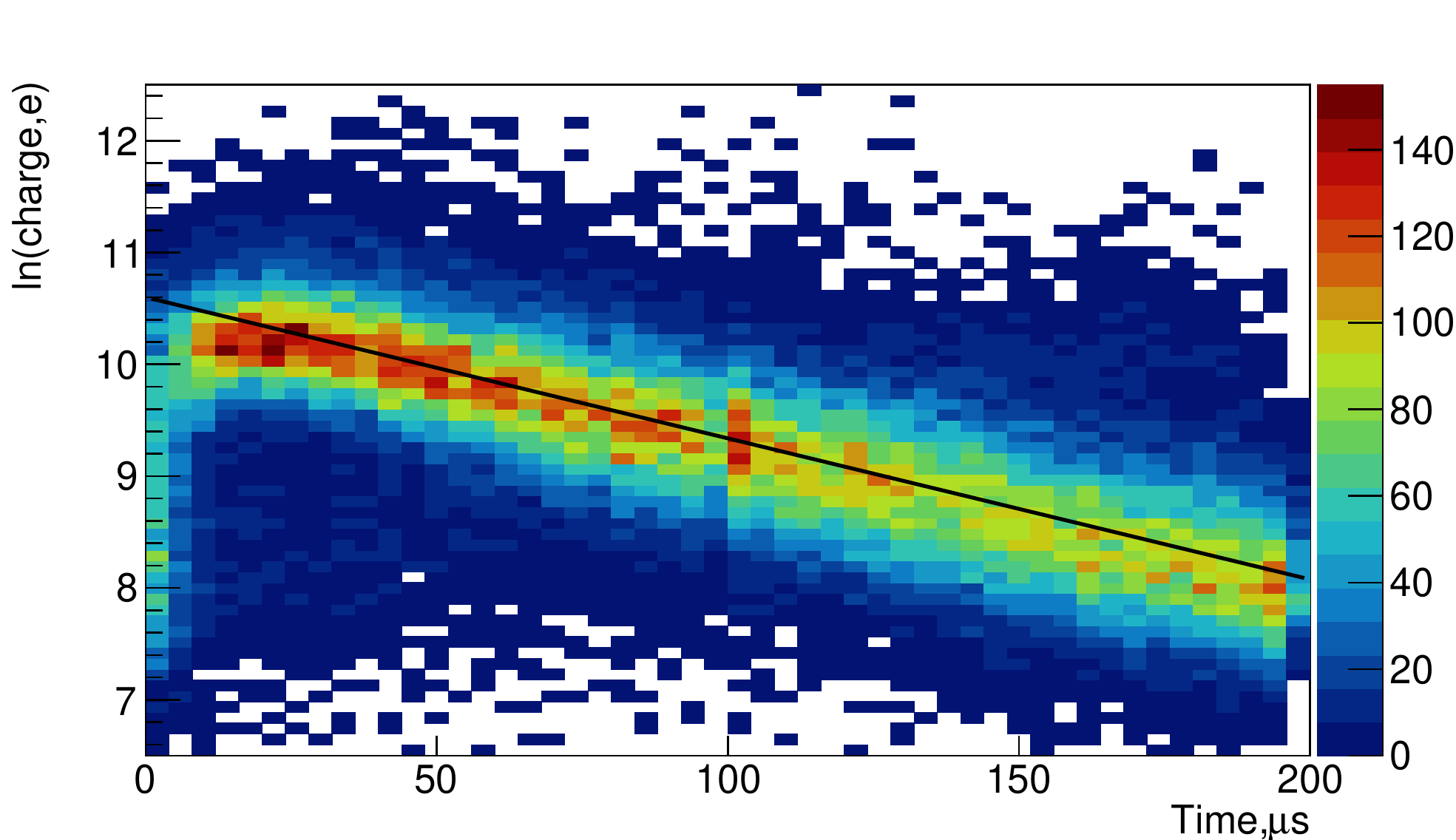}
        \caption{\centering U plane hit charge vs time}
        \label{fig:EliftimeU}
    \end{subfigure}\\

    \begin{subfigure}[b]{0.484\textwidth}
        \centering
        \includegraphics[width=0.95\textwidth]{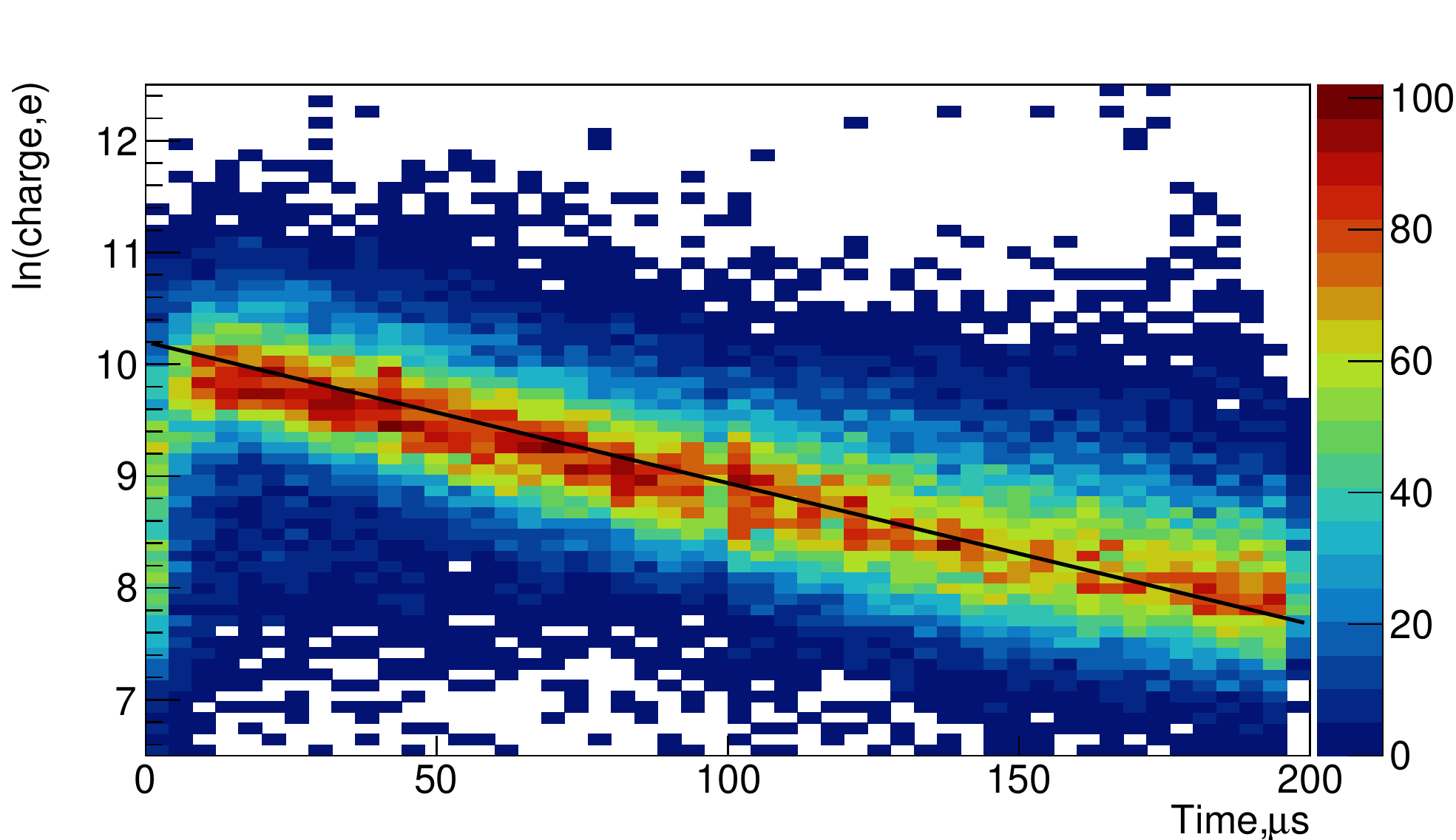}
        \caption{\centering V plane hit charge vs time}
        \label{fig:EliftimeV}
    \end{subfigure}
    
    \caption{The logarithm of the deposited charge vs time for selected long muon tracks for all three wire planes. The black solid line on each plot represents the average electron lifetime.}
    \label{fig:Eliftime}
\end{figure}

The logarithm of the deposited charge is plotted against the time and shown in Fig.~\ref{fig:Eliftime} for all three wire planes. The electron lifetime is derived from the profile linear fit for each wire plane separately. We achieved an average electron lifetime measurement among the three planes of 77~$\mu$s. It is shown as a solid line on all three plots. The achieved electron lifetime is sufficient for the measurement as follows from the detector response study, described next.

\subsection{Detector Response Uniformity}

We also present the study of the detector response, given that the uniformity of the response plays a crucial role in the beam attenuation measurement to extract the cross section. 

\begin{figure}
    \begin{subfigure}[b]{0.48\textwidth}
    \centering
        \includegraphics[width=1\textwidth]{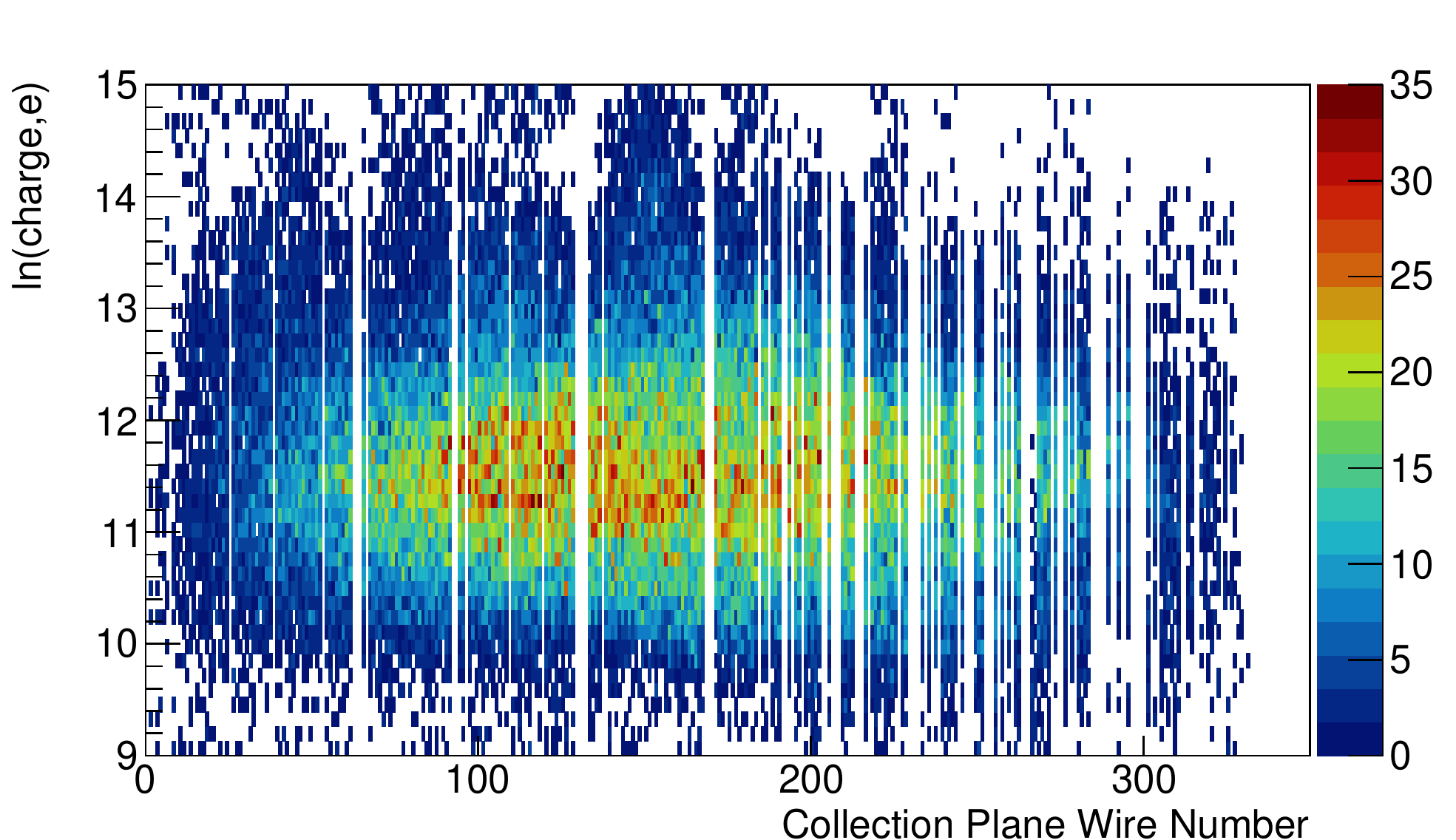}
        \caption{\centering X plane measured charge}

    \end{subfigure}
  \newline
    \begin{subfigure}[b]{0.48\textwidth}
    \centering
        \includegraphics[width=1\textwidth]{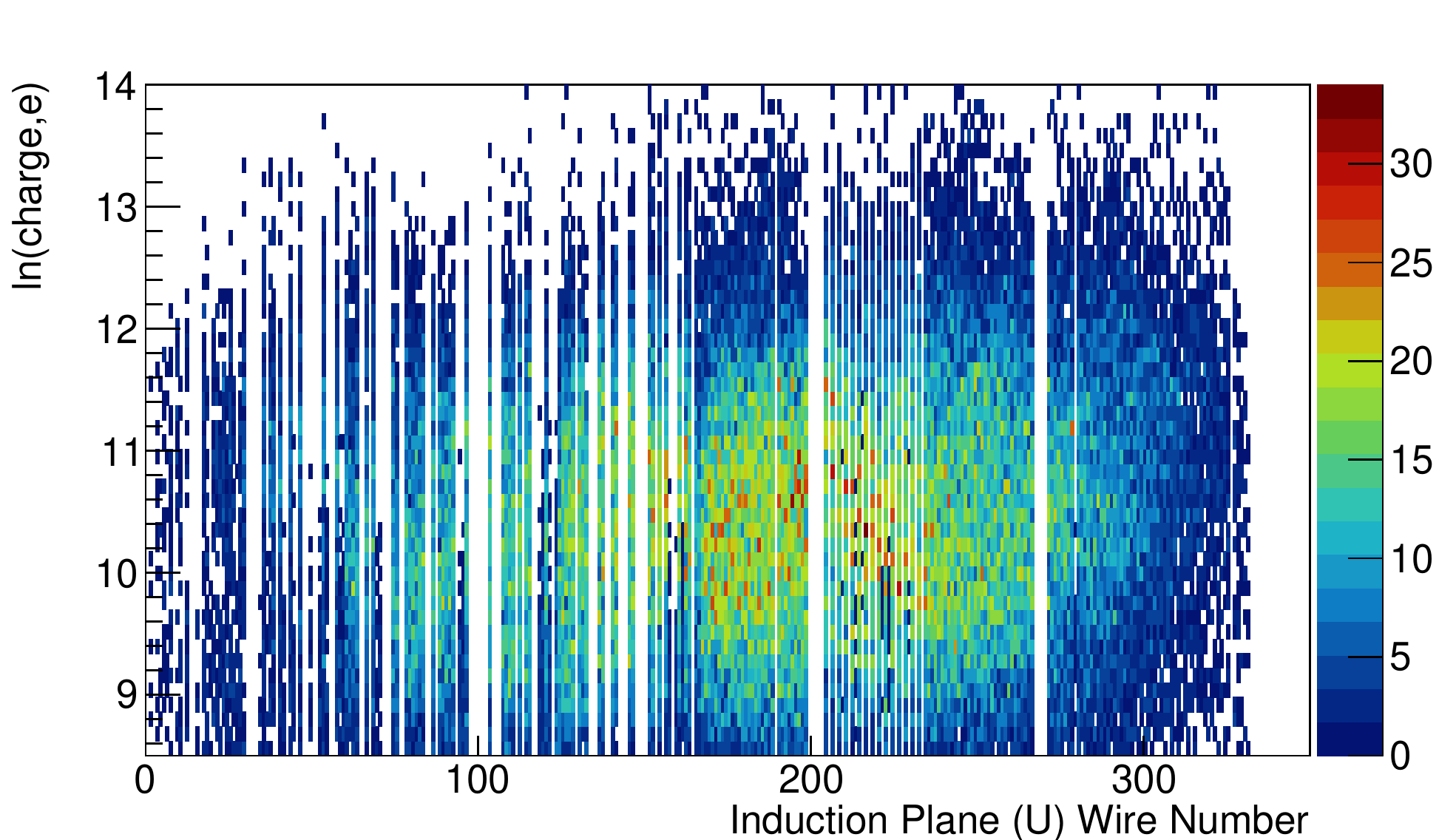}
        \caption{\centering U plane measured charge}
    \end{subfigure}
\newline
    \begin{subfigure}[b]{0.48\textwidth}
    \centering
        \includegraphics[width=1\textwidth]{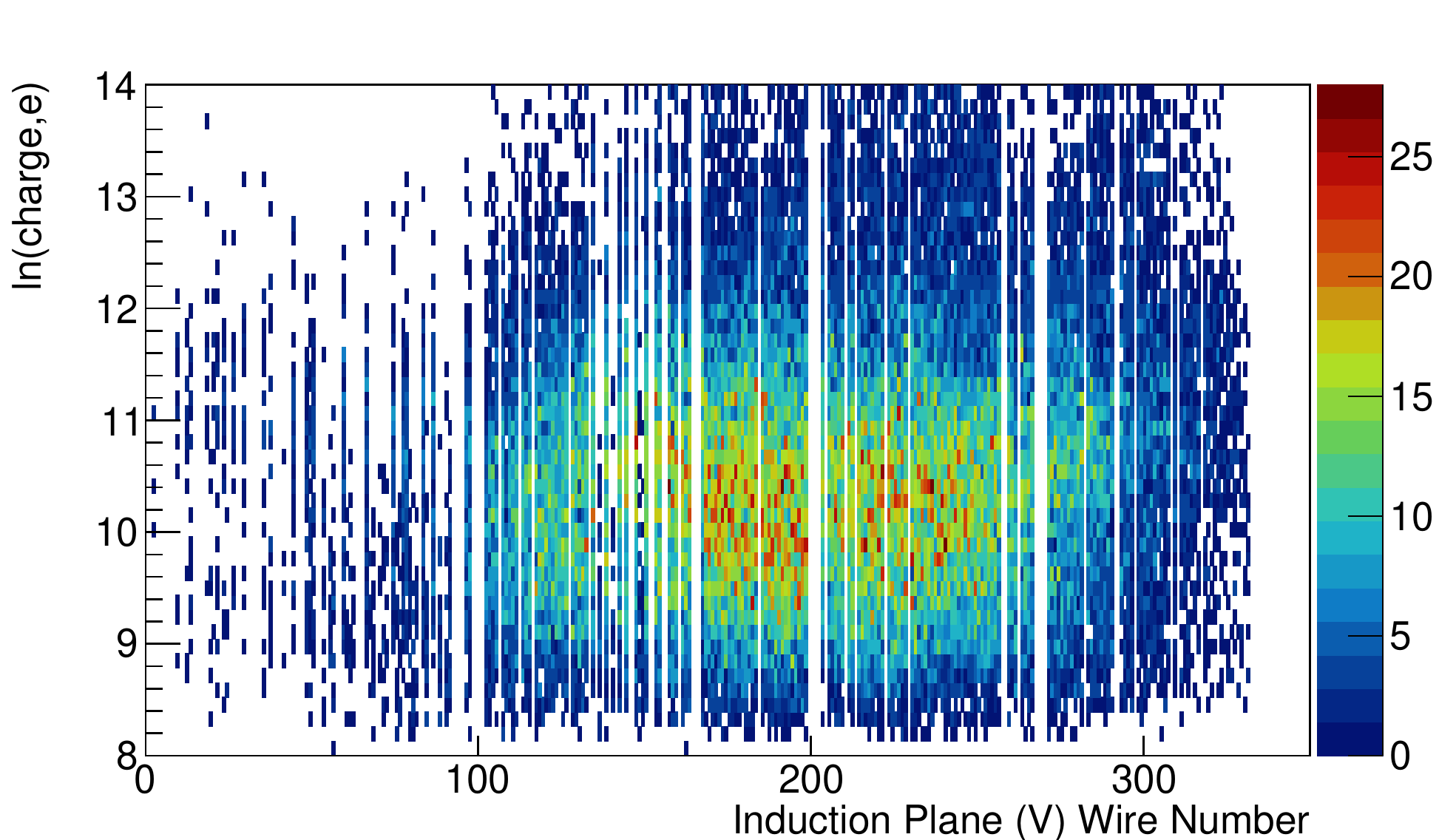}
        \caption{\centering V plane measured charge}
    \end{subfigure}
    \caption{Collected charge for each wire for all three wire planes. Blank regions represent wires either turned off during construction or eliminated as inefficient.}
    \label{fig:CollectedCharge}
\end{figure}

Figure \ref{fig:CollectedCharge} shows the measured charge for each wire for all three wire planes from the cosmic muon sample. Blank regions represent wires either turned off during construction or eliminated as inefficient. The remaining wires provide a calibration of the wire energy response for the Mini-CAPTAIN electronic simulation. 

The study shows that the mean measured charge is uniform across the detector, with more uninstrumented wires in the upstream part of the detector. Moreover, the measured charge from minimum ionizing particles (MIPs) is well above the chosen threshold (wires should measure at least 20 electrons to show signal).

The other study presents the wire efficiency as the neutron beam travels along the X-axis (decreasing wire number). We calculate efficiency only for a 50~mm radius region around the beam, since this region is used in the analysis. As previously discussed, each reconstructed track is split into three wire plane projections for the analysis. For each projection, we identify two reconstructed hits based on their timing information. One of these hits should be registered right before the beam time window, while the other one is right after. Next, based on the time difference between these two hits, the hit time is predicted for all wires in between. If a wire's predicted time falls in the beam window, the denominator in the efficiency calculation for this wire is incremented ($N_{tot}$ in Equation~\ref{equation:effWire}). If the considered wire has a reconstructed hit in the predicted time window the numerator in the efficiency calculation for this wire is incremented ($N_{app}$ in Equation~\ref{equation:effWire}). We define the efficiency for each instrumented wire in the detector as:

\begin{equation}
    \epsilon =\frac{N_{app}}{N_{tot}}
    \label{equation:effWire}
\end{equation}

\begin{figure}
    \begin{subfigure}[b]{0.48\textwidth}
    \centering
        \includegraphics[width=0.93\textwidth]{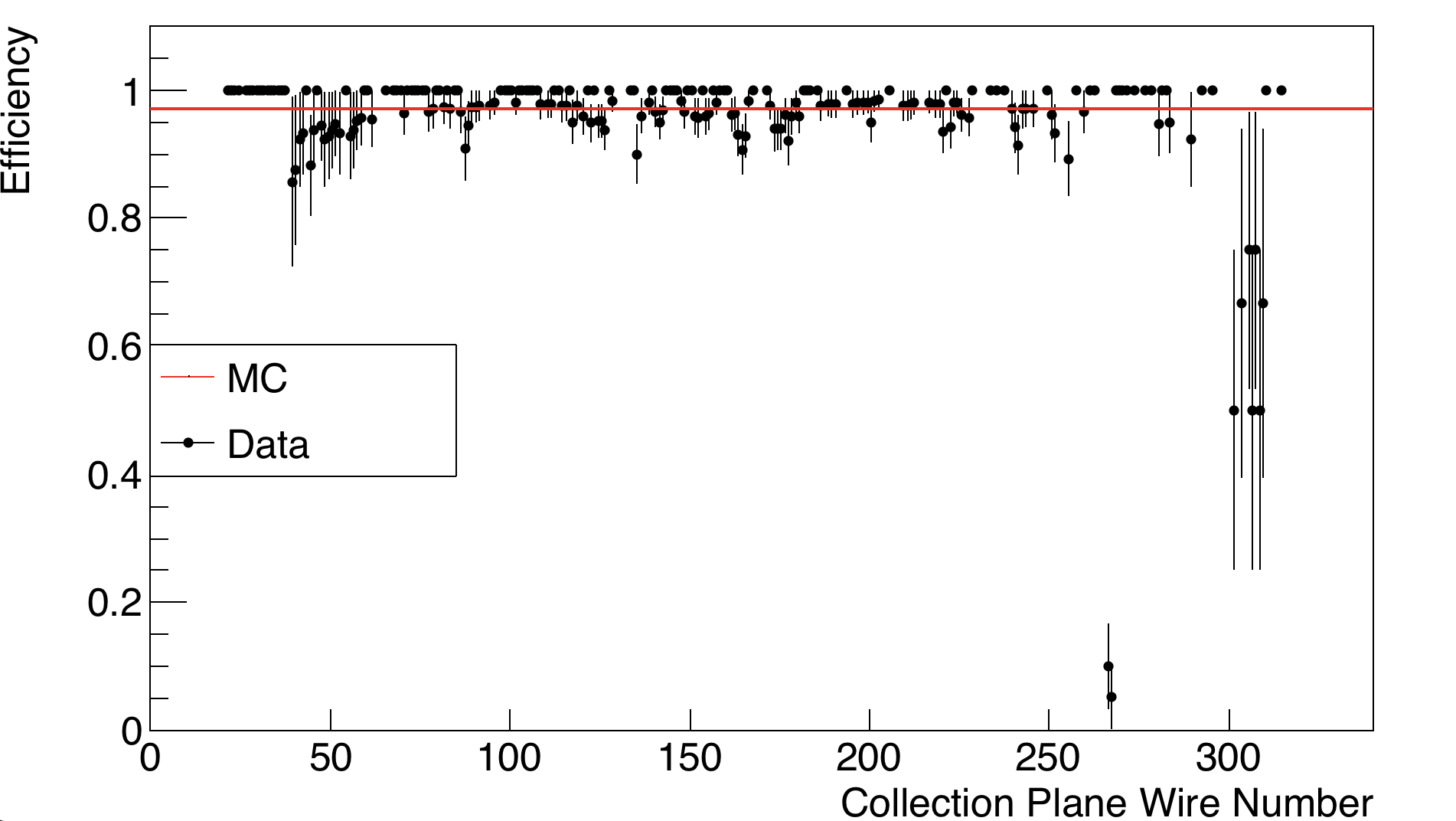}
        \caption{\centering X plane wire efficiency}

    \end{subfigure}
  \hfill
    \begin{subfigure}[b]{0.48\textwidth}
    \centering
        \includegraphics[width=0.93\textwidth]{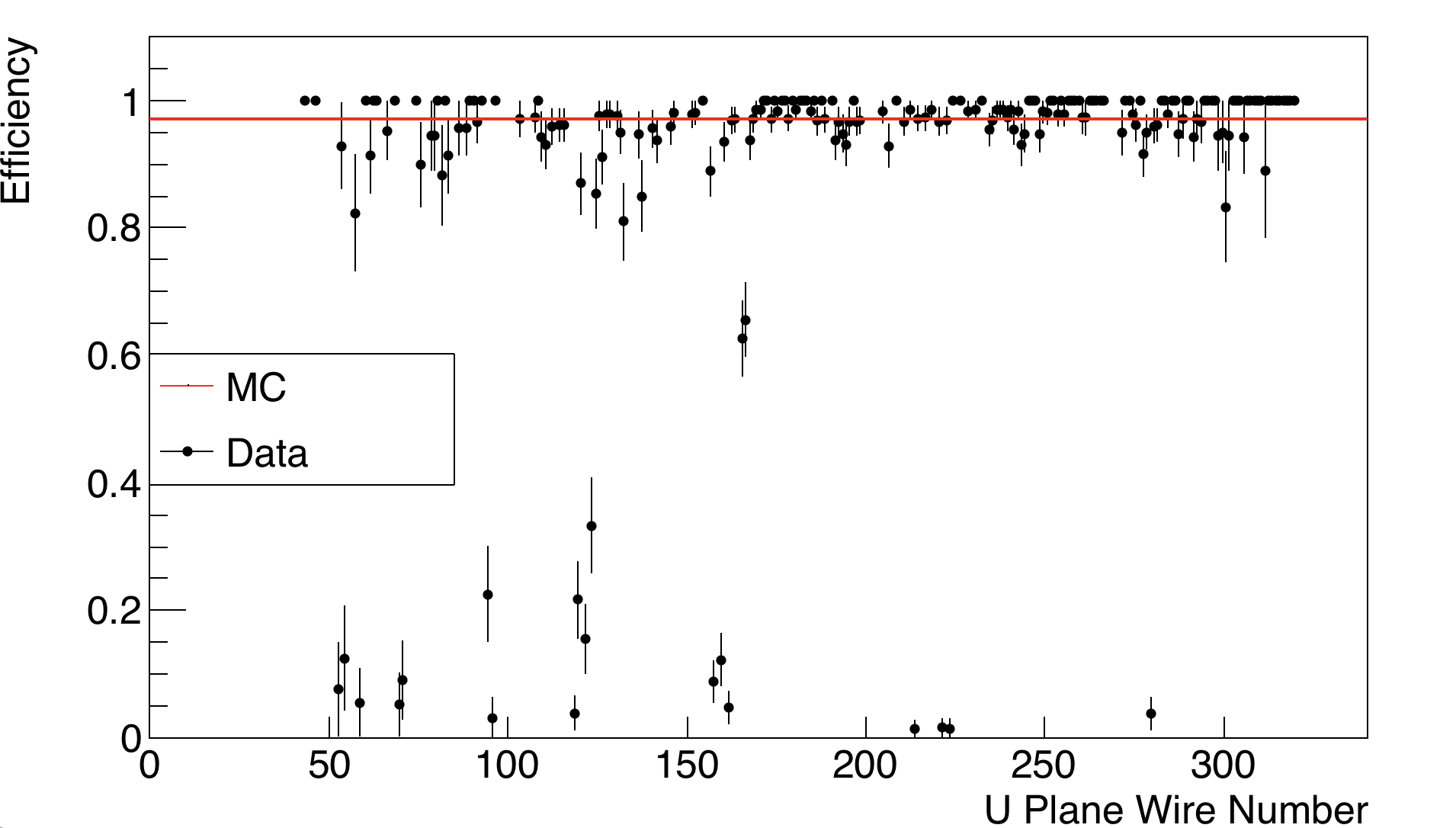}
        \caption{\centering U plane wire efficiency}
    \end{subfigure}
\newline
    \begin{subfigure}[b]{0.48\textwidth}
    \centering
        \includegraphics[width=0.93\textwidth]{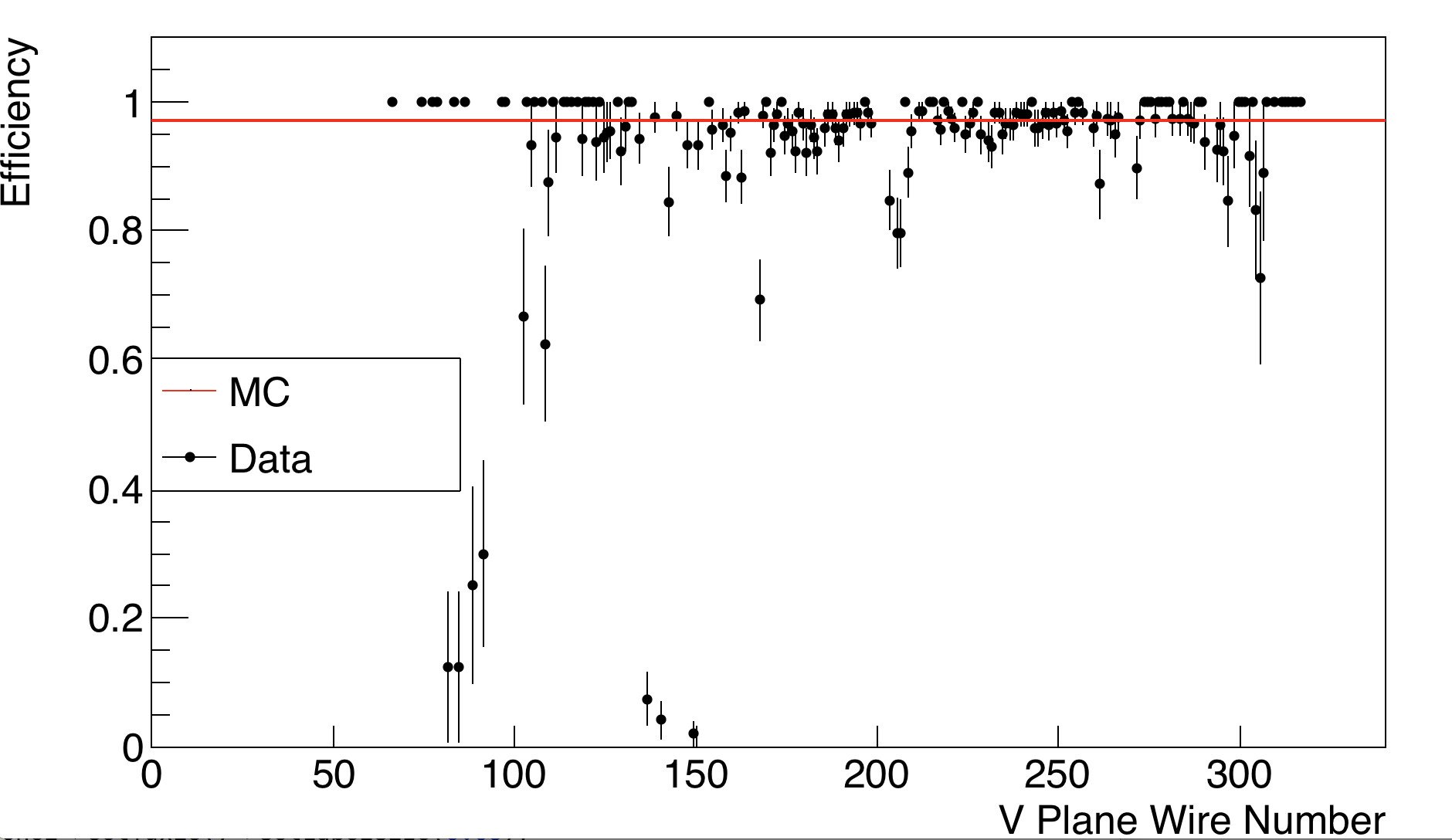}
        \caption{\centering V plane wire efficiency}
    \end{subfigure}
    \caption{Efficiency for each wire for all three wire planes in a 50~mm region around the beam. The black dots represent data, and the solid line is the nominal MC efficiency.}
    \label{fig:effWireDriftBeam}
\end{figure}

Figure \ref{fig:effWireDriftBeam} shows the result of the wire efficiency for each wire plane around the beam. Wires in the MC are simulated with the same efficiency of about 97\% around the beam spot, shown as a red line in the figure. The result shows that wire efficiency does not change across the detector with the exception of a few wires, thus providing the uniform response needed for the cross section measurement. Moreover, the simulated efficiency is in close agreement with the data. We consider the remaining difference in the wire efficiency between MC and data as a systematic uncertainty. 

\section{Simulation}
\label{sec:Sim}

\subsection{Beam Study}

We derive parameters for the simulated beam based on the beam shape observed in the experiment. The detector is oriented such that the beam enters it at the 4th quadrant with nearly maximum possible X coordinate (and wire number) and close to zero, but negative Y. Thus, we define the starting position of all tracks to be the one with the highest X coordinate.

\begin{figure}[h]
    \begin{subfigure}[b]{0.5\textwidth}
        \includegraphics[width=0.95\textwidth]{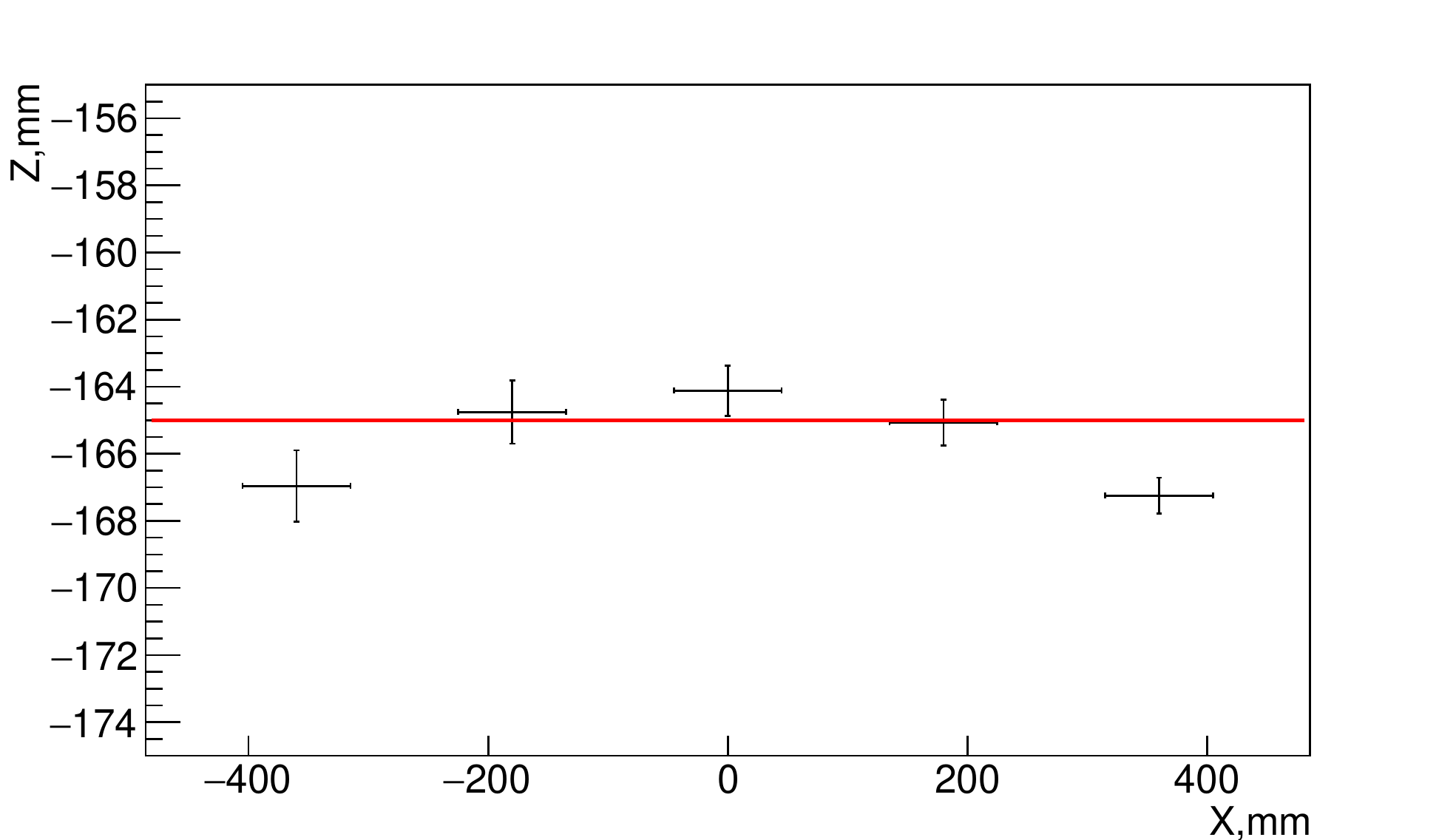}
        \caption{\centering Mean values }
    \end{subfigure}
  \hfill
    \begin{subfigure}[b]{0.5\textwidth}
        \includegraphics[width=0.95\textwidth]{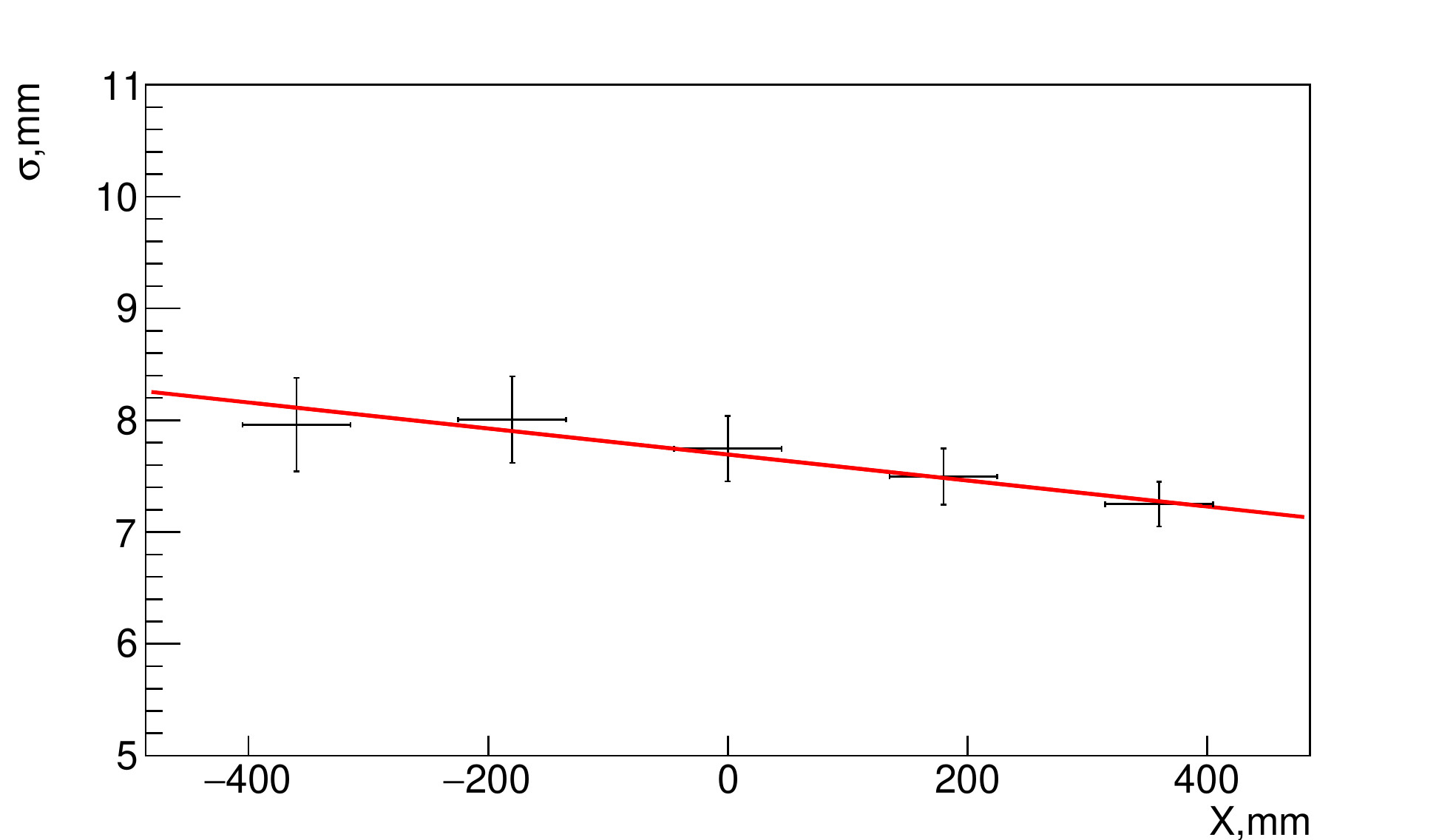}
        \caption{\centering Sigma values}
    \end{subfigure}

    \caption{Mean values (a) and sigma values (b) of Gaussian data peaks in Z for all five slices in X. The solid line is best fit line.}
    \label{fig:Zmean}
\end{figure}

\begin{figure}[h]
    \begin{subfigure}[b]{0.5\textwidth}
        \includegraphics[width=0.95\textwidth]{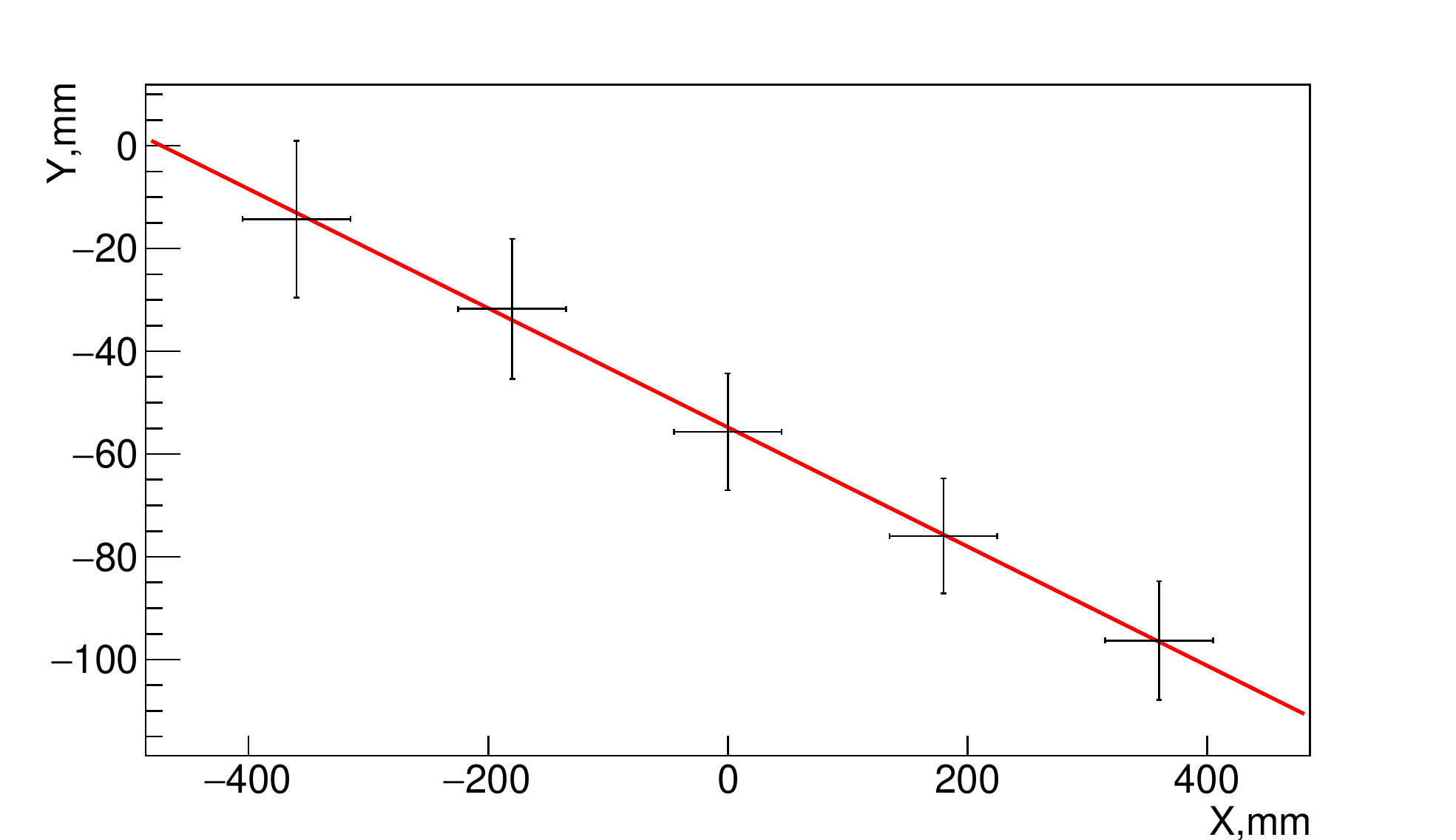}
        \caption{\centering Mean values}
    \end{subfigure}
  \hfill
    \begin{subfigure}[b]{0.5\textwidth}
        \includegraphics[width=0.95\textwidth]{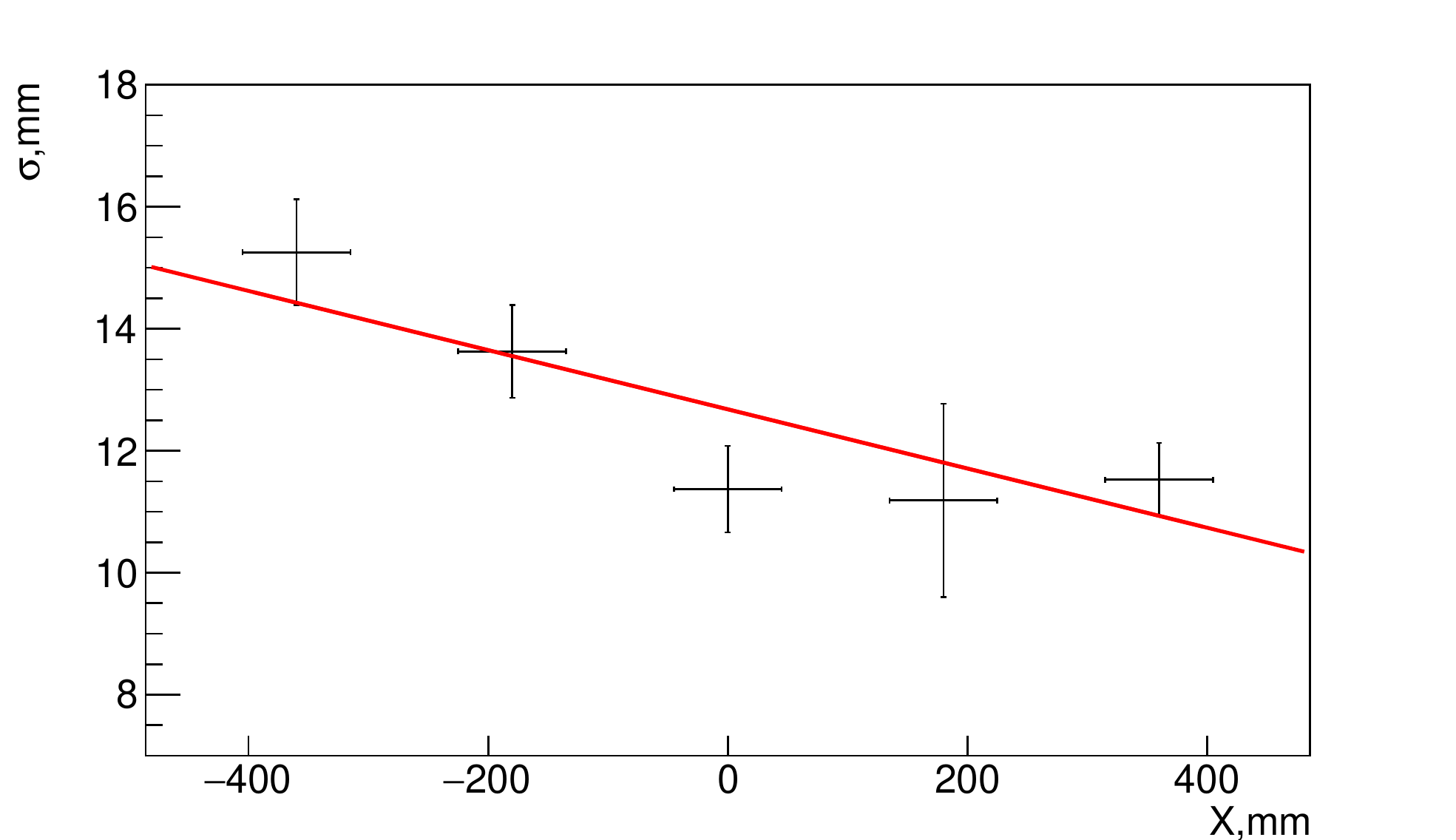}
        \caption{\centering Sigma values}
    \end{subfigure}

    \caption{Mean values (a) and sigma values (b) of Gaussian data peaks in Y for all five slices in X. The solid line is best fit line.}
    \label{fig:Ymean}
\end{figure}

The events with one reconstructed track can be used to study the shape of the beam. The detector is divided into 5 equal slices in X between -450~mm and 450~mm. We fit the distributions of starting Y and Z position of tracks for each slice. The summary of all mean values and sigma values from the Gaussian fits is shown in Fig.~\ref{fig:Zmean}(a) and \ref{fig:Zmean}(b) respectively for Z position and in Fig.~\ref{fig:Ymean}(a) and \ref{fig:Ymean}(b) for Y position. 
We fit the above distributions with a straight line to define the Y and Z properties of the beam as it travels through the detector. 

As a result, the beam travels parallel to the XY plane and with a $6.7$~degree angle to the X-axis. The beam spread in Y and Z at the cryostat boundary is 9.3~mm and 7~mm respectively. The beam center propagation is best described by the line:
\begin{equation}
    y=-0.1188\times x -54.9894,~z=-165~\rm{mm}
    \label{eq:beamLine}
\end{equation}

\subsection{Detector Simulation}

We use GEANT$\_$4.10.3 with the QGSP$\_$BERT physics list\cite{geant} to simulate the detector volume and the energy depositions by the charged particles. We also use the NEST \cite{Szydagis_2011} model to properly calculate the number of thermal electrons and scintillation photons. Ionization electrons are allowed to drift toward wire planes via the applied electric field. We describe the drift as an exponential model based on the observed electron lifetime with a small amount of diffusion added to the electron velocity and path. 

We use the Shockley-Ramo theorem to estimate induced current in wires after the cloud of ionization electrons reaches the wire planes.

The shape of signal is modeled by assuming an electron drifted directly to the collection wire plane. More specifically, the electron drifts in a straight line past the induction planes, reaches the collection plane, and travels to the closest collection wire. Wires that are not directly impacted by the electron are considered shielded. We consider this model sufficient since the cross section measurement is not based on precise calculations of energy deposition. All wires are used in a binary fashion, which means they have a signal or not based on the amount of charge received. The threshold choice is based on the  detector performance with MIPs from cosmic data. The uncertainty in the threshold is included in the systematic uncertainties.  

We model continuum and discreet noise separately. The continuum noise is modeled as $1/f^\alpha$ plus Gaussian noise and put directly in the impedance calculation. The non-continuum (discreet) noise has a relatively narrow bandwidth around a set of specific frequencies. We identify the positions of the noise peaks from the neutron data. The amplitude is chosen using a Gaussian with the mean power of the data peak and the phase is derived from a uniform distribution. We observe that the phase between the different frequencies is not correlated. 

Finally, we create a simulated set of neutron interactions. It includes 1.7 million events with one neutron per event. Neutrons start 23.2~m from the center of the cryostat and follow the best fit beam line given by Equation~\ref{eq:beamLine}. The spread of the beam in the Z and Y directions is simulated to be 24~mm to ensure enough statistics in the tails of the distribution.

\section{Data Analysis}
\label{sec:Data}

The final section focuses on the extension of the method of measuring the neutron cross section in liquid argon described in previous CAPTAIN paper~\cite{mypaper}. First, the fit structure is described. Second, the section focuses on fit validation and systematic studies. Finally, the fit application to the neutron data is described as well as the final cross section results. 

\subsection{Event selection}

We select events for the analysis based on the same criteria for the data and MC:
\begin{itemize}
    \item Only one reconstructed track in the total analysis region (50~mm radius around the beam center) per micropulse for the data and per event for MC;
    \item Reconstructed track should be at least 15~mm long in the X projection (beam direction) and start in fiducial volume (between -400~mm and 400~mm in X).
\end{itemize}

We separate all events based on neutron time-of-flight (TOF). The TOF bins for the analysis are chosen based on PDS information from the neutron data, the 4~ns resolution of the PDS system, and available statistics for each TOF range. In total, the we operate with five TOF bins:  [140-180]~ns, [120-140]~ns, [112-120]~ns, [104-112]~ns, [96-104]~ns.

\begin{figure}
\begin{center}
\includegraphics[width=0.45\textwidth]{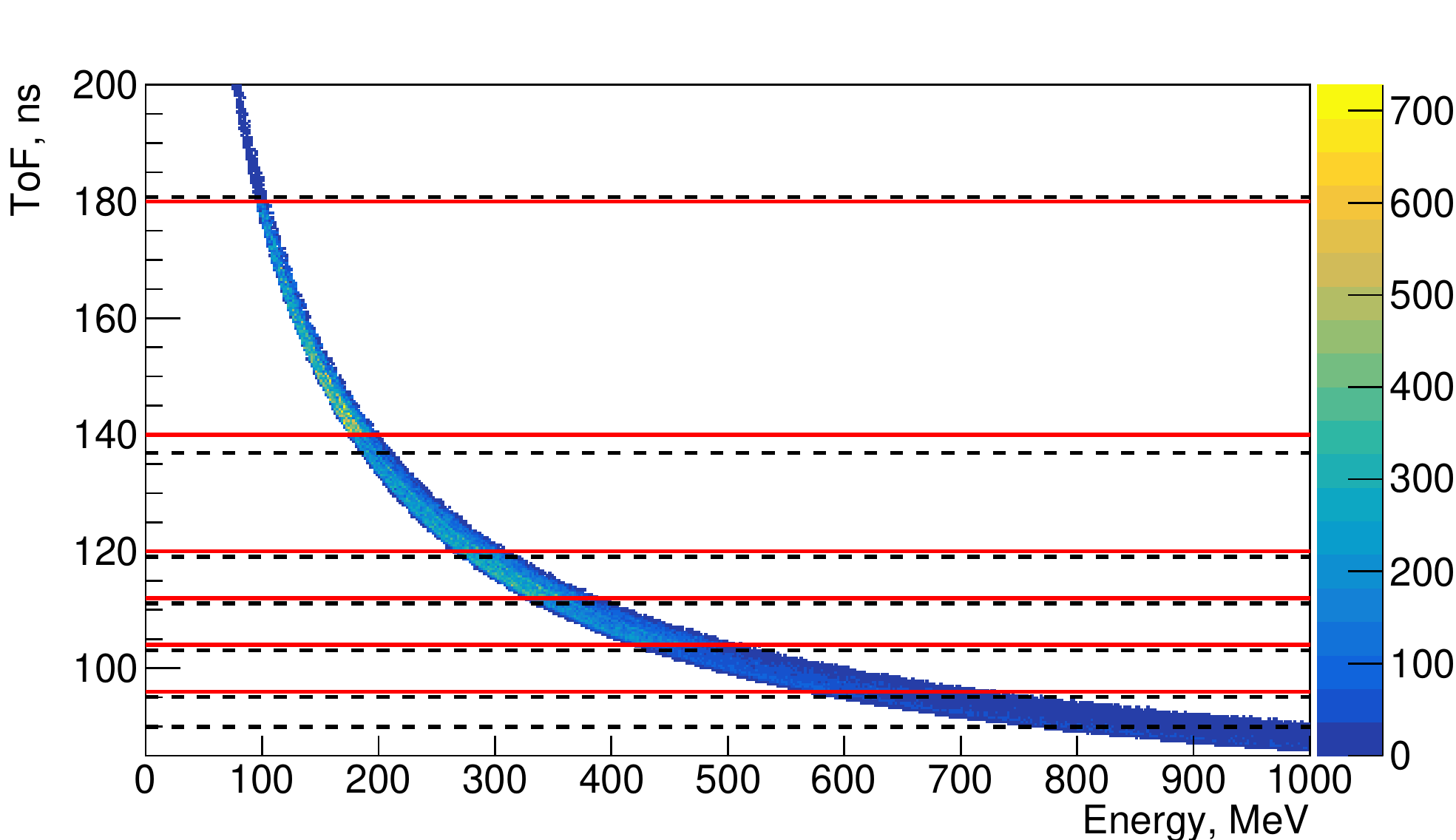}
\end{center}
\caption{Neutron TOF calculated based on MC simulation vs initial neutron energy. The solid red lines are the chosen TOF bins in this measurement. The black dashed lines are the energy bins used in the previous measurement.}
\label{fig:TOFBINS_mc}
\end{figure}

The neutron TOF for the MC simulation is calculated based on the initial neutron energy and the distance between the source and the first ``visible" interaction inside the detector. We call the interaction ``visible" if it produces a charged particle with a track longer than 10~mm inside the detector. Figure~\ref{fig:TOFBINS_mc} shows the selected TOF bins and corresponding neutron energy ranges in between solid lines. Dashed lines represent the energy bins previously used\cite{mypaper}. The neutron energy ranges corresponding to the selected TOF bins with flux averaged energies are listed in Table~\ref{tab:TOFBINS_mc}. The translation between the neutron energy and time-of-flight is non-linear. Thus, we conclude that the switch to TOF bins instead of energy bins is essential to ensure that each event corresponds to a unique bin. We cut the energy at 720~MeV because the addition of an extra TOF bin would include events from the unphysical region (with reconstructed energy above 800~MeV), which can't be reconciled with 800 MeV neutrons given the uncertainties of the PDS. 

\begin{table}[h]
\centering
\caption{The neutron energy range and the flux weighted average energy for each TOF bin according to MC simulation using the neutron energy spectrum provided by LANSCE\cite{nowicki2017alamos}.}
\begin{tabular}{ c | c | c }
\hline\hline
\makecell{TOF range, [ns]} & \makecell{Energy Range, [MeV]}  & \makecell{Flux averaged \\ energy, [MeV]}\\
\hline\hline
140-180 & 95-200 & 143\\ 
\hline
120-140 & 174-315 &  236\\ 
\hline
112-120 & 265-385  & 319\\ 
\hline
104-112 & 325-515  & 404\\ 
\hline
96-104 & 420-720 &  543\\ 
\hline\hline
\end{tabular}

\label{tab:TOFBINS_mc}
\end{table}

We split the data into two regions for each TOF bin based on the most upstream position of the reconstructed track measured along the beam axis, the signal and the side regions. The signal region includes reconstructed tracks starting within a 25~mm radius around the best fit beam center, and the side region includes tracks  with starting positions within 25-50~mm radius around the best fit beam center. We also chose the fiducial volume to cover the full detector drift region excluding 50~mm margins on each side, which covers the distance between -400~mm and 400~mm along the X-axis. 

The total number of considered events in the data (using the same neutron data as before) is 5810, which is 2.4 times more than used in the initial CAPTAIN measurement \cite{mypaper}. The statistics available in each TOF bin are presented in Table~\ref{tab:TOFBINS_stat}.

\begin{table}[h]
\centering
\caption{Number of events in each TOF bin for the neutron data.}
\begin{tabular}{ c | c }
\hline\hline
\makecell{TOF range, [ns]} & \makecell{Number Of Events}  \\
\hline\hline
140-180 & 625  \\ 
\hline
120-140 & 1344   \\ 
\hline
112-120 & 985   \\ 
\hline
104-112 & 1272   \\ 
\hline
96-104 & 1584   \\ 
\hline\hline
\end{tabular}

\label{tab:TOFBINS_stat}
\end{table}

\subsection{Fitting function and algorithms}

Events in data and MC are divided into 50 total bins ($n_{bins}$), consisting of ten bins in the X coordinate and five TOF bins.  Of the ten bins in X, six are in the signal region and four are in the side region, described in the previous section.  We fit for the cross section by minimizing the following function:

\begin{equation}
\label{eq:ch1}
    \chi^2=\sum_{i=1}^{n_{bins}}\frac{Data_i-MC_i(\sigma,\alpha,\delta)}{Data_i}+C(\alpha,\delta)
\end{equation}
where $Data_i$ is the number of data events in bin $i$, $MC_i$ is the re-weighted number of MC events in bin $i$, C is a function that constrains the fitting parameters, and $\sigma$, $\alpha$, and $\delta$ are the fitting parameters. Each event in the MC simulation is re-weighted based on the starting position of the initial neutron and the event topology.

In order to properly re-weight the MC, we first modify the initial shape of the beam. We assume that the beam has a Gaussian shape in both Y and Z directions. The beam spread (sigma) observed in data and interpolated to the detector entrance is about 9.3~mm in the Y direction and 7~mm in the Z direction. The simulation uses a spread of 24~mm in each direction. Thus, each MC event is assigned an initial weight based on the starting point of the initial neutron. The simulated distribution of initial neutrons is fitted with function:
\begin{equation}
    g=p_0e^{-\frac{(x-p_1)^2}{2p_2^2}}
\end{equation}
Next, parameter $p_2$ is set to mimic the beam spread observed in the data. The final beam shape weight assigned to each MC event is the following:
\begin{equation}
    \label{eq:W_beamshape}
    W_{Beam Shape}=\frac{g_{y_{new}}(y_n)}{g_{y_{initial}}(y_n)}\times\frac{g_{z_{new}}(z_n)}{g_{z_{initial}}(z_n)}
\end{equation}
where g is Gaussian fit with (new) and without (initial) the corrected $p_2$ parameter in both directions, and $y_n$ and $z_n$ are the corresponding coordinates of the initial neutron.

Next, we re-weight each MC event based on the event topology. We define all event topologies based on ``visible" neutron interactions in the detector. There are three event categories used in the analysis. The first is the ``Signal" category, which is defined as the following:
\begin{itemize}
    \item The true track from the first ``visible" interaction has an initial neutron as a parent particle;
    \item The true track starting position deviates less than 0.1~mm from the path of the initial neutron.
\end{itemize}
The second category is called ``Elastic" and represents an event with a neutron undergoing one or multiple elastic scatterings prior to the ``visible" interaction. Moreover, interactions with low energy transfer which do not change the ID of the particle in the GEANT simulation are also a part of this category. Thus, the category is defined as following:
\begin{itemize}
    \item The true track from the first ``visible" interaction has an initial neutron as a parent particle;
    \item The true track starting position deviates more than 0.1 mm from the path of the initial neutron.
\end{itemize}
The final category for the re-weighting is called ``Other". It combines all other possible event topologies including inelastic scattering, gamma production, etc. Since the events are binned based on the information coming from the reconstruction, some MC events do not have any true information associated with them. These events fall into a separate category called ``NoTrueInfo". These events are weighted for the beam spread but not the event topology.

We assign a weight for each category based on the interaction probability of the neutron inside the detector. The probability is given by the equation:

\begin{equation}
    P_{\rm surv}=e^{-T\times l\times\sigma^{\rm tot}}
\end{equation}
where $\sigma^{\rm tot}$ is the total neutron cross section, $l$ is the distance the neutron traveled in the medium, and  $T=\rho_{\rm LAr}\times N_{\rm Avogadro}/m_{\rm Ar}$ is the nuclear density of liquid argon. The nuclear density is a constant in the experiment and equals $2.11\times10^{22}$~cm$^{-3}$ ($T=(1.3973$~g/cm$^3\times6.022\times10^{23}$~neutrons/mol$)/39.948$~g/mol). Thus, each ``Signal" category event can be assigned a weight of:
\begin{equation}
    W_{\rm Signal}=\frac{e^{-T\times l\times\sigma^{\rm tot}_{\rm new}(\rm TOF)}}{e^{-T\times l\times\sigma^{\rm tot}_{\rm MC}(\rm TOF)}}
\end{equation}
where the $\sigma^{\rm tot}_{\rm new}(\rm TOF)$ represents the fitted cross section parameter for a given TOF bin. The $\sigma^{\rm tot}_{\rm MC}(\rm TOF)$ represents the base flux averaged GEANT value for the neutron cross section for a given TOF bin.

We obtain values of $\sigma^{\rm tot}_{\rm MC}(\rm TOF)$ for each TOF bin using a simulated thin target measurement. We simulate two million neutrons with energy equal to the fluxed averaged energy of a given TOF bin. The large number of events allowed for a negligible statistical uncertainty. The cross section is extracted using the attenuation of the beam after 1~cm travel distance. The neutron is considered to have interacted if a charged particle track with a length above 15~mm in the X projection is produced. The results are presented in Table~\ref{tab:base_cs}. 

\begin{table}[h]
\centering
\caption{The cross section values used as a base values in the fit ($\sigma^{tot}_{MC}(TOF)$). Values obtained via simulated thin target measurement for each TOF bin.}
\begin{tabular}{ c | c }
\hline\hline
\makecell{TOF range, [ns]} & \makecell{Base cross section \\ value, [b]}  \\
\hline\hline
140-180 & 0.59  \\ 
\hline
120-140 & 0.53   \\ 
\hline
112-120 & 0.53   \\ 
\hline
104-112 & 0.56   \\ 
\hline
96-104 & 0.58   \\ 
\hline\hline
\end{tabular}

\label{tab:base_cs}
\end{table}

The weight for the ``Elastic" category events is assigned in a similar way with an additional parameter describing the branching ratio between two event categories:
\begin{equation}
\label{eq:elast}
    W_{\rm Elastic}=\frac{e^{-T\times l\times\sigma^{\rm tot}_{\rm new}(\rm TOF)}}{e^{-T\times l\times\sigma^{\rm tot}_{\rm MC}(\rm TOF)}}\times e^{-\alpha(\rm TOF)}
\end{equation}
where the $\alpha(\rm TOF)$ represents the fitting branching ratio parameter for a given TOF bin. The ``Elastic" category can include multiple elastic events prior to the main interaction as well as other effects. Thus, the $\alpha$ parameter accounts for all these effects. The exponent prevents the weight from being negative. The base value of the $\alpha$ parameter is zero.

We define the final ``Other" category weight as a free parameter:

\begin{equation}
\label{eq:other}
    W_{\rm Other}=e^{-\delta(\rm TOF)}
\end{equation}
where the $\delta(\rm TOF)$ represents the fitting parameter for a given TOF bin. The exponent prevents the weight from being negative. The base value of the $\delta$ parameter is zero. 

Finally, we do the MC normalization separately for each TOF bin. The numerator of the normalization coefficient($N_{\rm Data}$) is the total number of neutron data events in a given TOF bin (both signal and side regions). The denominator($N_{\rm MC}$) is the total number of re-weighted MC events in a given TOF bin. The final form of the normalization coefficient is given as:
\begin{equation}
\label{eq:norm}
    \eta(\rm TOF)=\frac{N_{\rm Data}(\rm TOF)}{N_{\rm MC}(\rm TOF)}
\end{equation}

The combination of all re-weighting steps gives the final form of the term $MC_i$ in the initial $\chi^2$, Equation~\ref{eq:ch1}:
\begin{equation}
    MC_i=\eta(\rm TOF)\times\sum_{j=1}^{N_i}(W_{\rm BeamShape}\times W_{\rm Category})
\end{equation}
where $\eta(\rm TOF)$ is given by EquationQ\ref{eq:norm} for a given TOF bin, the $N_i$ is a total number of MC events in bin $i$, and $W_{\rm BeamShape}$ and $W_{\rm Category}$ are weights assigned for each event in the beam based on the starting position of the initial neutron and event topology respectively.

To summarise, the $\chi^2$ function given by Equation~\ref{eq:ch1} has five cross section parameters, one for each TOF bin. The ``Elastic" topology has five extra re-weighting parameters, one for each TOF bin ($\alpha$). The ``Other" topology has five parameters as well ($\delta$). These extra parameters for the ``Elastic" and the ``Other" topologies ($\alpha$ and $\delta$) are called topology parameters. The total number of bins to compare across all data is 50 (five TOF bins with six signal region bins and four side region bins in each TOF bin). Thus, the number of degrees of freedom in the problem is 35. The fit is performed using the ROOT implementation of MINUIT.

\subsection{Fit study}

We evaluate the fit stability by using a specifically simulated data set. The data set represents an average expectation of the experiment given the true cross section and topology parameters. These parameters can be set to a different values to evaluate the fit performance.

At first, all parameters are set to their base GEANT values given in Table~\ref{tab:base_cs}. Next, we apply the fitting technique described in the previous section to check the function behavior and fit convergence. After a preliminary study we observed numeric convergence problems because of high correlations between cross section and topology parameters in each TOF bin. Thus, we studied the $\chi^2$ function around the minimum, which showed the presence of a plateau region that might interfere with numerical minimization algorithms for all topology parameters. In order to fix this issue, we introduced a loose constraint for each topology parameter. The prior expectation of the variation of the exponential terms in equations \ref{eq:elast} and \ref{eq:other} is much less than the order of magnitude. Thus, we allowed the topology parameters to vary by $\pm$2.3, which allowed the exponential terms to vary by $\pm$10 around the base value. This variation is big enough not to interfere with the result of the fit, but sufficient to fix the fit convergence issue corresponding to the function plateau.

The application of the described constraint significantly improved the correlations. However, for the first TOF bin (largest time-of-flight or smallest energy neutrons) the correlations remain too high for the algorithm to perform a proper error calculation around the minimum. Since the energy of incoming neutrons in this TOF bin is low, we expect the number events in the ``Other" category to be low. Thus, the parameter $\delta_1$ is fixed to the default GEANT value (zero) and is excluded from the fit. These changes led to the stable numeric convergence of the given algorithms with proper second derivatives around the minimum. Moreover, the number of parameters is reduced by one which brings the number of degrees of freedom in the problem to 36. The constraint term in the Equation~\ref{eq:ch1} is given as:
\begin{equation}
    C=\sum_{k=1}^{5}\frac{\alpha_k^2}{2.3^2}+\sum_{k=2}^{5}\frac{\delta_k^2}{2.3^2}
\end{equation}
where $k$ represents the number of the TOF bin.

We perform the fit in its final form multiple times with various starting points for the cross section parameters to ensure that the algorithm finds the unique minimum. The fit successfully converged to the base cross section values (Table~\ref{tab:base_cs}) with a maximum divergence of 0.37\%.  

There are four major systematic effects on the cross section parameters that we studied:
\begin{itemize}
    \item The effect of ``Other" topology events;
    \item The effect of ``Elastic" events;
    \item The effect of  multiple-track events in the cross
section calculation;
    \item The effect of dead wires in the upstream region of the detector.
\end{itemize}

Since the ``Other" category includes primarily inelastic events, the parameters describing this category are closely related to the inelastic neutron cross section in argon. Thus, the maximum uncertainty that can be put on these parameters can be derived from the difference between the cross section measurement described in \cite{old_neutron} (for neutrons below 50~MeV) and the initial measurement by the CAPTAIN collaboration described in \cite{mypaper} (for energies above 100~MeV). The two experiments have different setups, energy ranges, and sensitivity. The inelastic threshold in liquid argon as given in \cite{old_neutron} is 1.5~MeV. However, the minimal detected energy loss that corresponds to signal interactions according to simulation in the Mini-CAPTAIN detector is 60~MeV. Taking all of this into account, we compare the cross section result at 50~MeV from \cite{old_neutron} against the CAPTAIN result at 100~MeV from \cite{mypaper}. 

These two results are different by as much as a factor of four. Thus, we determine the systematic uncertainty from the ``Other" category by changing the number of ``Other" events in the simulated data to four times its base value. The result for each TOF bin is presented in second column of the Table~\ref{tab:allfit}.  According to the study, only the first TOF bin has a significant systematic effect evaluated at 13.2\%. This is expected because of the $\delta_1$ parameter being fixed.

The evaluation of the systematic effect of ``Elastic" events on the cross section parameters is based on the same logic as described above. We set the maximum variation on the number of ``Elastic" events in the simulated data set to four times its base value. The result is presented in third column of the Table~\ref{tab:allfit}. The result suggests that fit is not sensitive to the given change in the ``Elastic" topology.

The effect of multiple-track events was the dominant systematic effect for the initial CAPTAIN cross section measurement. According to the simulation, the sample of multiple-track events in the selected region differs from the sample with only one track by a maximum of 5\%. Thus, we vary the total number of events in the simulated data set by $\pm$5\% in order to determine the effect of these events on cross section parameters. The forth column of the Table~\ref{tab:allfit} shows that the change in cross section does not exceed 2\%. Thus, the effect of multiple-track events is negligible in the new measurement. This is expected, since the inclusion of the upstream part of the detector in the analysis significantly improved signal selection.

The final systematic that we studied is caused by the differences in the wire efficiency in data and MC, in particular in the upstream part of the detector. All wires in the MC simulation have the same efficiency of 97\%, while actual wire efficiency varies between wires. The full comparison is presented in Fig.~\ref{fig:effWireDriftBeam}. The total number of active wires in the upstream part of the detector is 100 out of 165. In order to estimate the uncertainty on this number, the effective number of expected inefficient wires is calculated. The efficiency of each wire observed in the experiment in the upstream part of the detector is subtracted from the simulated efficiency. The sum of absolute values of these differences is 6.4 for the X plane. We use this number as a desired uncertainty. In order to study the effect of this uncertainty on the cross section measurement, we set each functional wire in the upstream part of the detector to be skipped in reconstruction with a 7\% probability. This led to an approximately 7-wire variation out of 100 simulated functional wires. The result presented in last column of the Table~\ref{tab:allfit} suggests that this is not a dominant systematic uncertainty in the experiment.

\begin{table}[h]
\centering
\caption{The summary of estimated systematic effects on cross section measurement.}
\begin{tabular}{ c | c | c | c | c }
\hline\hline
\makecell{TOF \\range, \\~[ns]} & \makecell{''Other" \\events, \\~[$\%$]}  & \makecell{''Elastic" \\events, \\~[$\%$]} & \makecell{Multiple-track \\events, \\~[$\%$]} & \makecell{Inefficient \\wires, \\~[$\%$]} \\
\hline\hline
140-180 & 13.2 & 0.34 & 0.339 & 3.05  \\ 
\hline
120-140 & 0.5 & 0.75 & 0.377 & 2.08   \\ 
\hline
112-120 & 1.8 & 1.5 & 0.377 & 0.57   \\ 
\hline
104-112 & 1.4 & 3.57 & 1.96 & 0.36   \\ 
\hline
96-104 & 1 & 1.89  & 0.862 & 0.34  \\ 
\hline\hline
\end{tabular}
\label{tab:allfit}
\end{table}

To summarize, we found that the dominant systematic uncertainty for the first TOF bin is caused by variation in the cross section of ``Other" events and is evaluated to be 13.2\%. For the rest of the TOF bins studied, systematic uncertainties do not exceed 4\%.

\subsection{Neutron data fit}

We use the finalized fitting procedure to study the neutron data. First, we run the fit multiple times on neutron data with various starting points for each cross section parameter to prove the existence of a unique minimum. The fit converged successfully to the same point with a maximum of 5\% variation across all cross section parameters. 

Second, we set all cross section parameters to the base GEANT values defined in Table~\ref{tab:base_cs} and all topology parameters to zero. The fit is performed using the ROOT implementation of MINUIT with the MINOS \cite{MINOS} extension to calculate parameter errors around the minimum. The cross section result of the fit for each TOF bin is presented in Table~\ref{tab:final_CS} with global correlations derived from the Hessian matrix. The results for ``Elastic" and ``Other" parameters are shown in Tables~\ref{tab:final_Elast} and \ref{tab:final_Other} respectively.

\begin{table}[h]
\centering
\caption{The post-fit cross sections for flux averaged energies inside each neutron TOF bin.  The Statistical error is calculated using the MINOS\cite{MINOS} algorithm.}
\begin{tabular}{ c | c | c | c  }
\hline\hline
\makecell{TOF range,\\~[ns]} & \makecell{Parameter}  & \makecell{Post-fit\\ cross section \\value, [b] } & \makecell{Global \\ correlation} \\
\hline\hline
\makecell{\\140-180 } & $\sigma(146~MeV)$ & $0.601^{+0.140}_{-0.143}$ & 0.111  \\ 
\hline
\makecell{\\120-140} & $\sigma(236~MeV)$ & $0.722^{+0.103}_{-0.101}$ & 0.138  \\ 
\hline
\makecell{\\112-120} & $\sigma(319~MeV)$ & $0.804^{+0.129}_{-0.121}$ & 0.226   \\ 
\hline
\makecell{\\104-112} & $\sigma(404~MeV)$ & $0.739^{+0.135}_{-0.091}$ & 0.544   \\ 
\hline
\makecell{\\96-104} & $\sigma(543~MeV)$ & $0.741^{+0.088}_{-0.088}$ & 0.429   \\ 
\hline\hline
\end{tabular}
\label{tab:final_CS}
\end{table}

\begin{table}[h]
\centering
\caption{The post-fit ``Elastic" category parameter values for each neutron TOF. The statistical error is calculated using the Hessian matrix. }
\begin{tabular}{ c | c | c | c | c  }
\hline\hline
\makecell{TOF \\ range,\\~[ns]} & \makecell{Parameter}  & \makecell{Post-fit \\ parameter\\ value} & \makecell{Statistical \\ uncertainty} & \makecell{Global \\ correlation} \\
\hline\hline
140-180 & $\alpha_1$ & $-4.3\times10^{-3}$ & 0.219 & 0.111  \\ 
\hline
120-140 & $\alpha_2$ & 0.176 & 0.539 & 0.915  \\ 
\hline
112-120 & $\alpha_3$ & 0.143 & 0.887 & 0.929  \\ 
\hline
104-112 & $\alpha_4$ & 0.283 & 0.771 & 0.892   \\ 
\hline
96-104 & $\alpha_5$ & -0.211 & 0.208 & 0.528  \\ 
\hline\hline
\end{tabular}
\label{tab:final_Elast}
\end{table}

\begin{table}[h]
\centering
\caption{The post-fit ``Other" category parameter values for each neutron TOF. The statistical error is calculated using the Hessian matrix. }
\begin{tabular}{ c | c | c | c | c }
\hline\hline
\makecell{TOF \\range, \\~[ns]} & \makecell{Parameter}  & \makecell{Post-fit \\ parameter \\ value} & \makecell{Statistical \\ uncertainty} & \makecell{Global \\ correlation} \\
\hline\hline
140-180 & - & - & - & -  \\ 
\hline
120-140 & $\delta_2$ & $-1.01\times10^{-3}$ & 0.606 & 0.915  \\ 
\hline
112-120 & $\delta_3$ & -0.035 & 0.713 & 0.931   \\ 
\hline
104-112 & $\delta_4$ & 0.029 & 0.679 & 0.898   \\ 
\hline
96-104 & $\delta_5$ & 2.051 & 1.238 & 0.373   \\ 
\hline\hline
\end{tabular}
\label{tab:final_Other}
\end{table}

\begin{figure*}
    \begin{subfigure}[b]{0.45\textwidth}
        \includegraphics[width=1.0\textwidth]{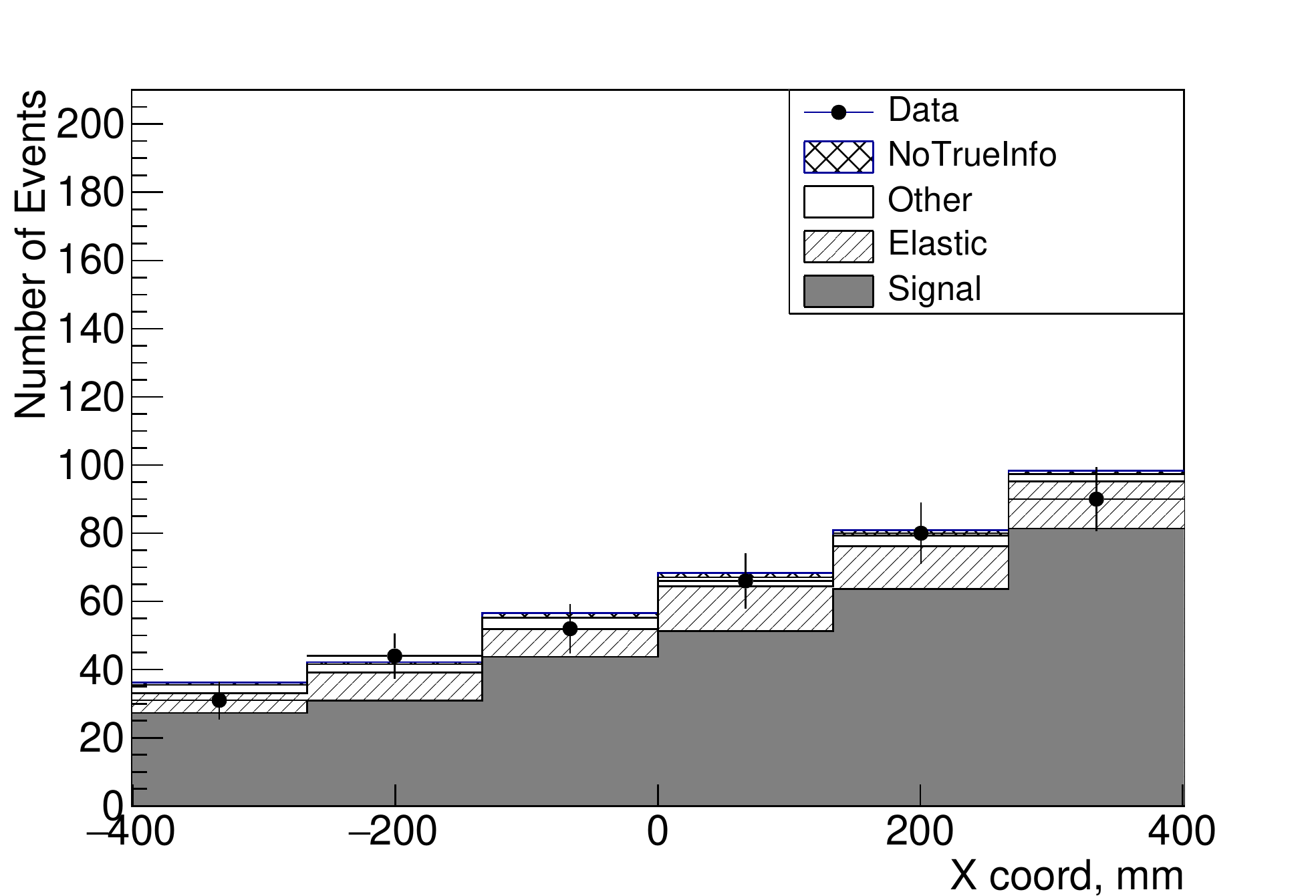}
        \centering
        \caption{\centering TOF=[140,180]ns \\ $\chi^2/NDF=2.23/6$ }
    \end{subfigure}
  \hfill
    \begin{subfigure}[b]{0.45\textwidth}
        \includegraphics[width=1.0\textwidth]{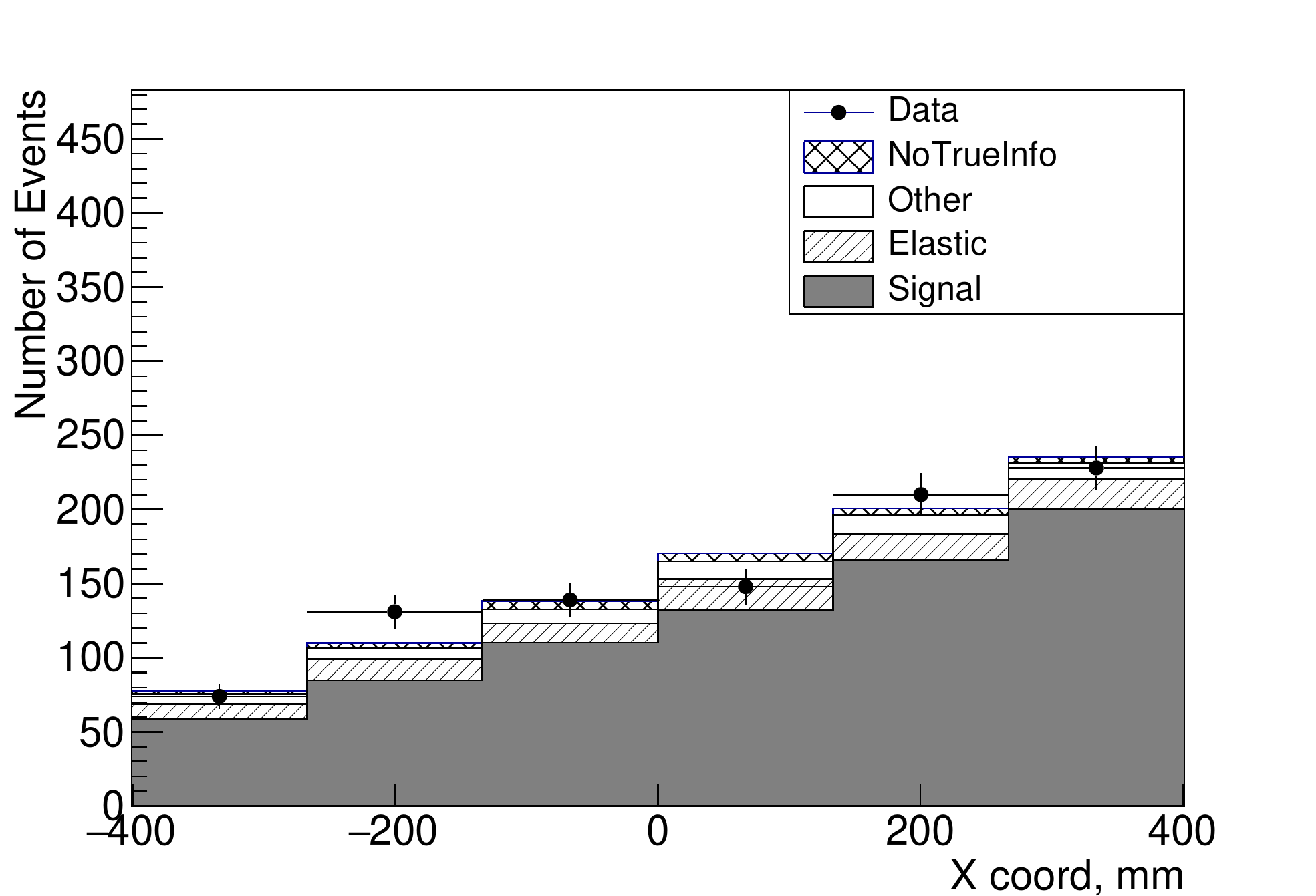}
        \caption{\centering TOF=[120,140]ns\\ $\chi^2/NDF=7.65/6$}
    \end{subfigure}
  \newline
    \begin{subfigure}[b]{0.45\textwidth}
        \includegraphics[width=1.0\textwidth]{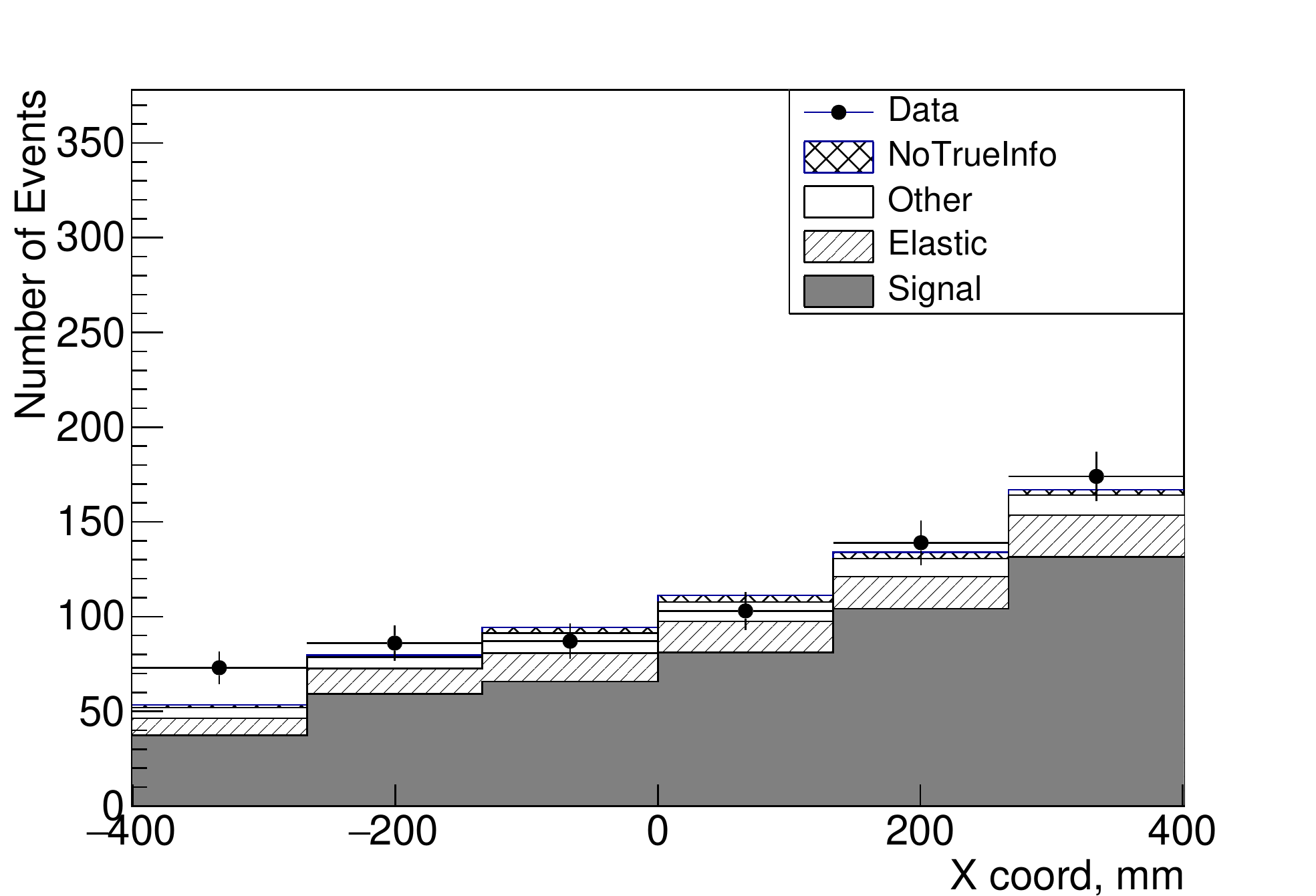}
        \caption{\centering TOF=[112,120]ns\\ $\chi^2/NDF=7.48/6$}
    \end{subfigure}
    \hfill
    \begin{subfigure}[b]{0.45\textwidth}
        \includegraphics[width=1.0\textwidth]{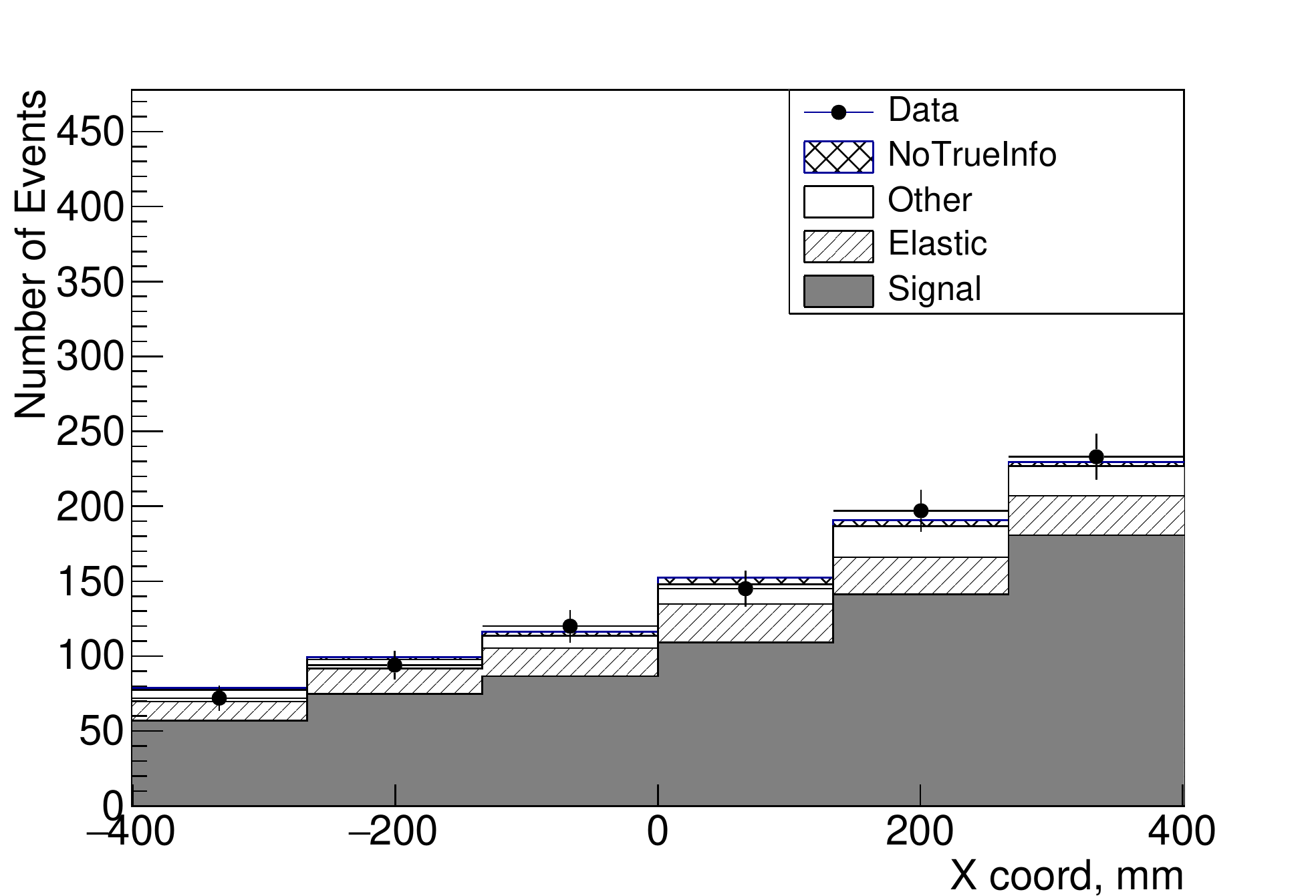}
        \caption{\centering TOF=[104,112]ns\\ $\chi^2/NDF=1.716/6$}
    \end{subfigure}
    \newline
      \begin{subfigure}[b]{0.5\textwidth}
      \centering
        \includegraphics[width=1\textwidth]{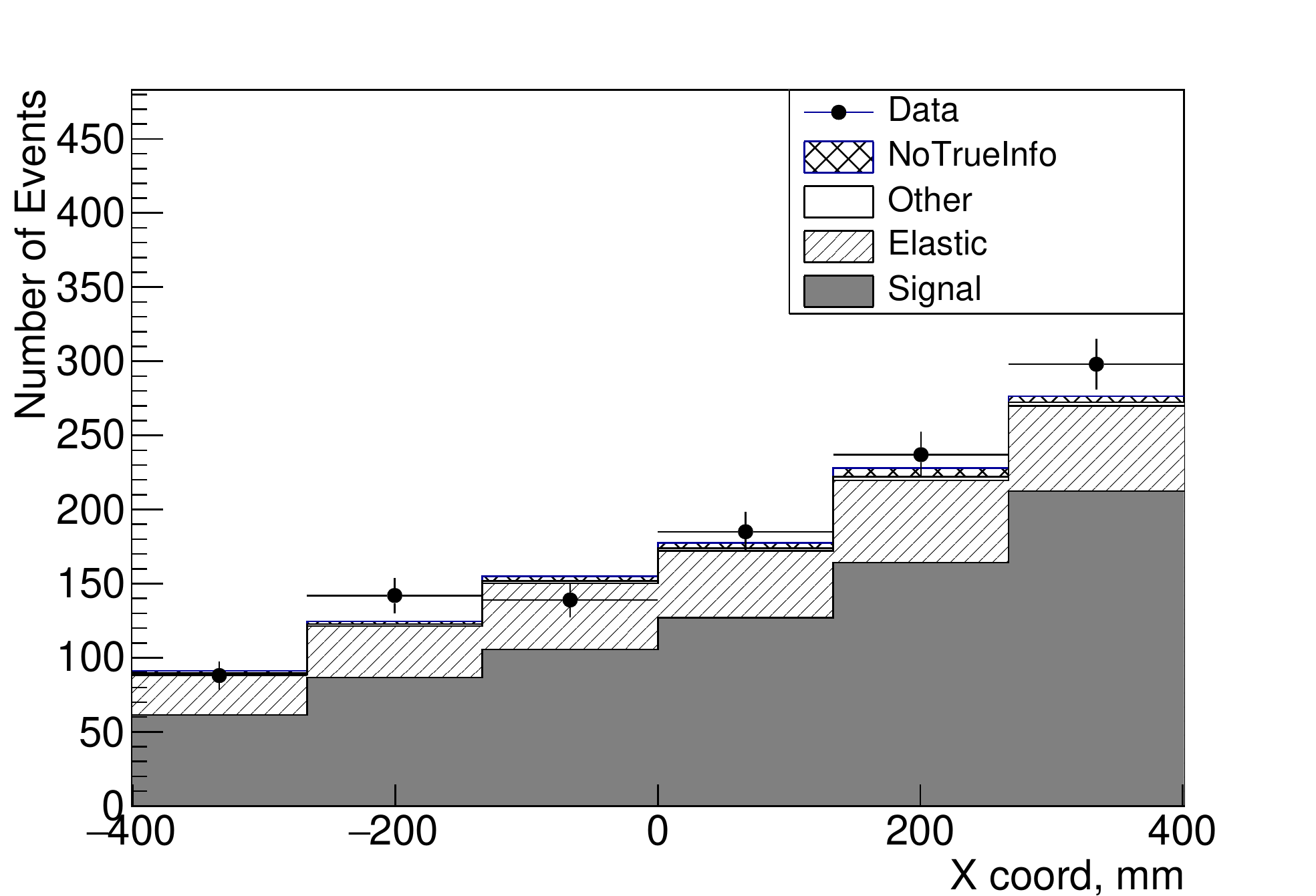}
        \caption{\centering TOF=[96,104]ns\\ $\chi^2/NDF=6.33/6$}
    \end{subfigure}
    
    \caption{Post-fit distribution of starting position of reconstructed tracks inside the signal region for each TOF bin.}
    \label{fig:TOF_compare_sig}
\end{figure*}

\begin{figure*}
    \begin{subfigure}[b]{0.45\textwidth}
        \includegraphics[width=1.0\textwidth]{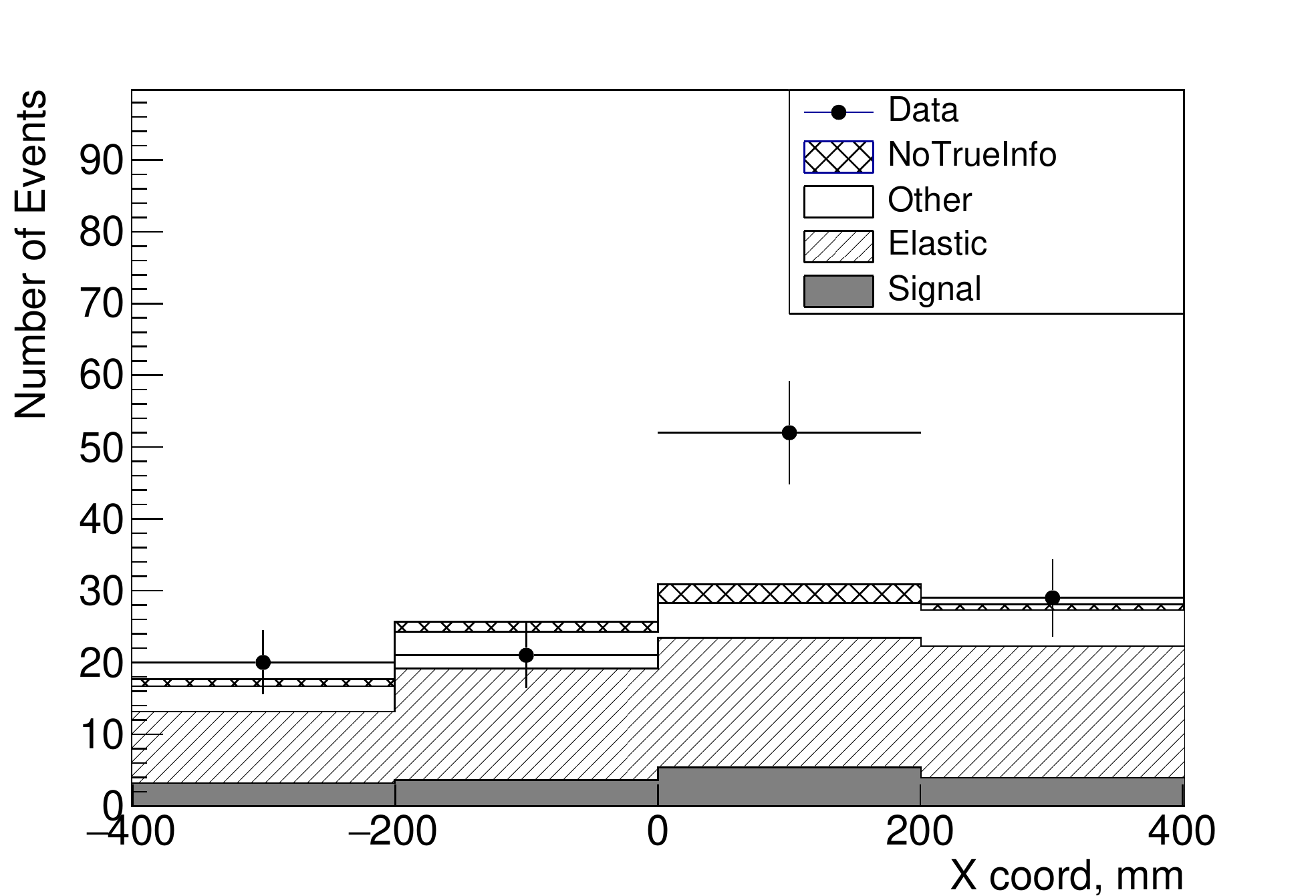}
        \caption{\centering TOF=[140,180]ns\\ $\chi^2/NDF=9.897/4$}
    \end{subfigure}
  \hfill
    \begin{subfigure}[b]{0.45\textwidth}
        \includegraphics[width=1.0\textwidth]{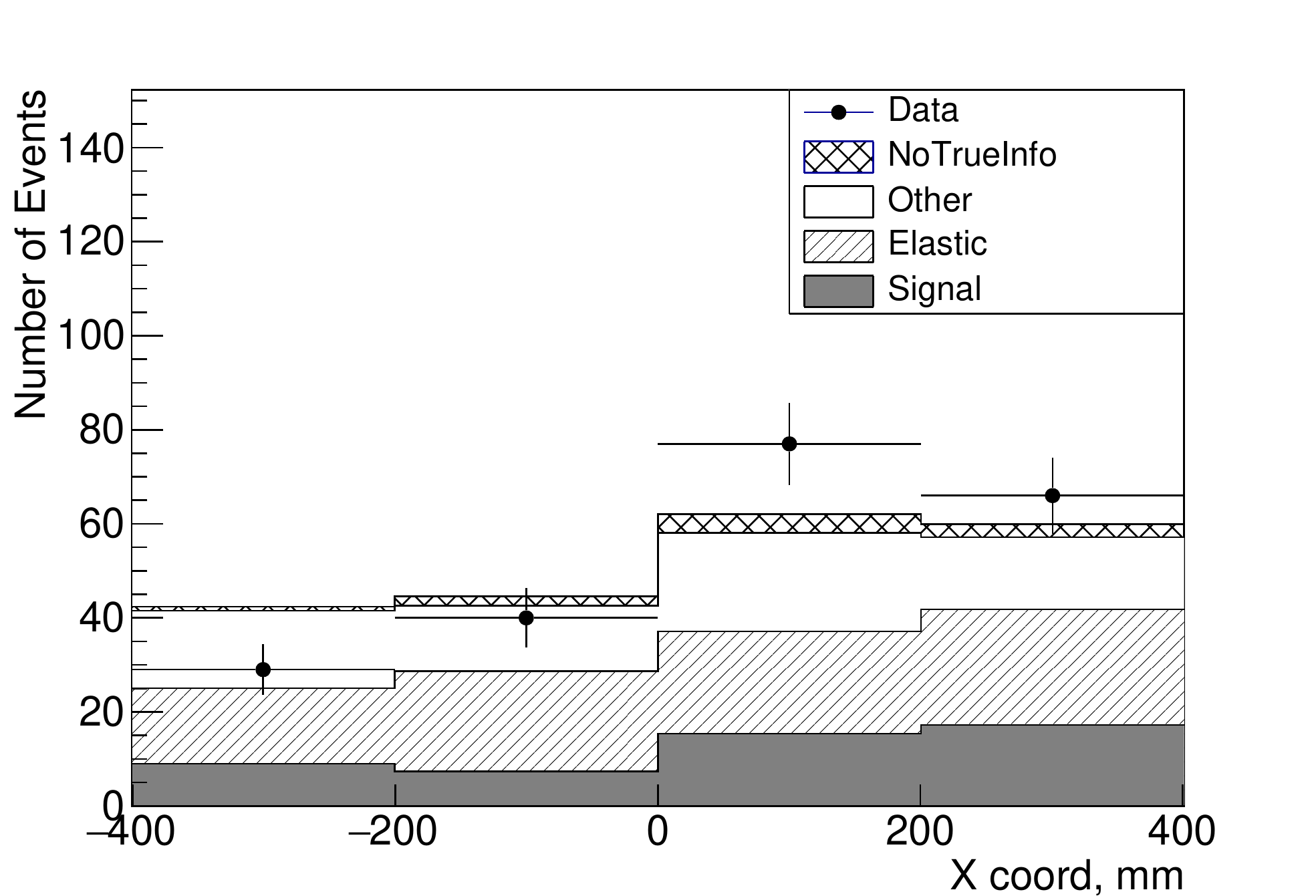}
        \caption{\centering TOF=[120,140]ns\\ $\chi^2/NDF=10.16/4$}
    \end{subfigure}
  \newline
    \begin{subfigure}[b]{0.45\textwidth}
        \includegraphics[width=1.0\textwidth]{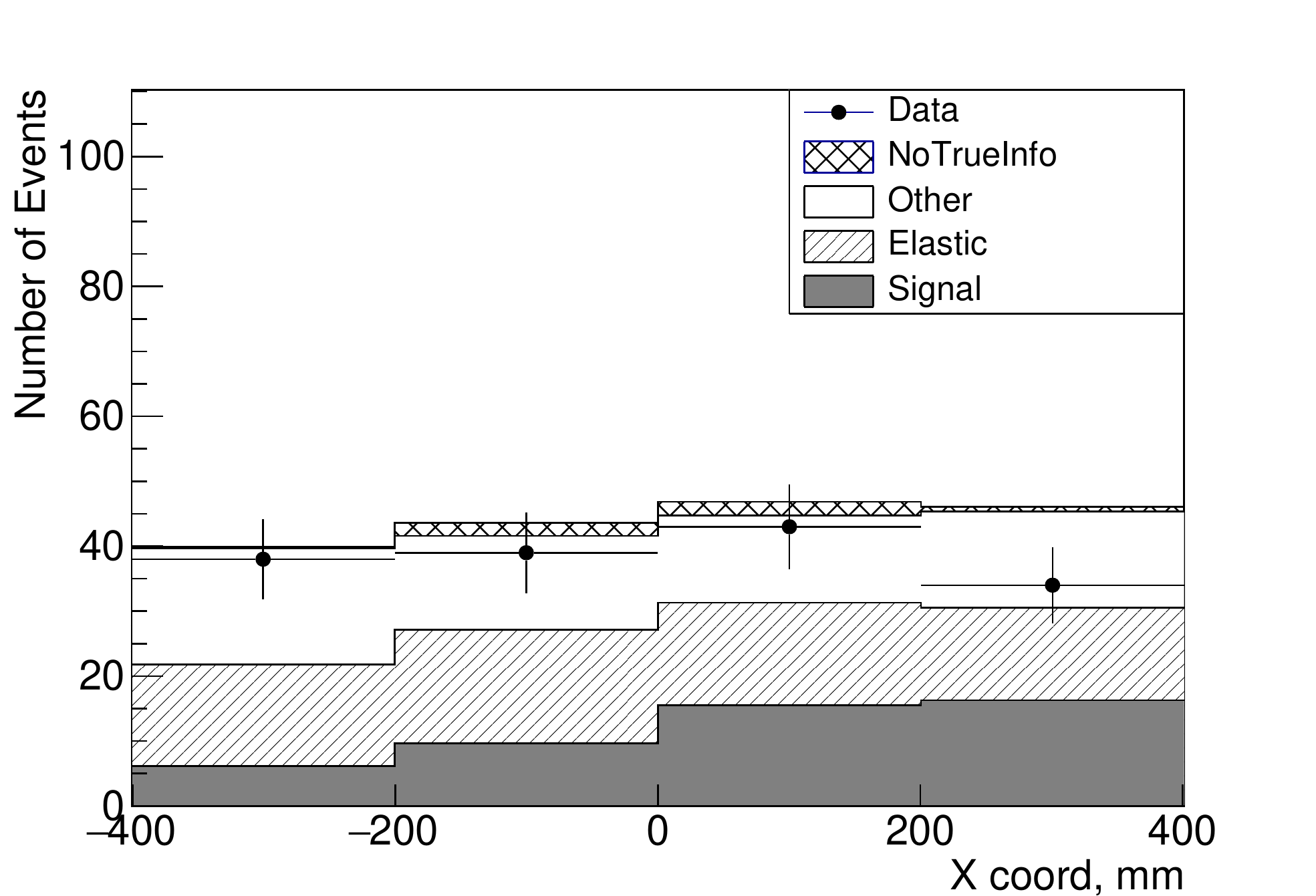}
        \caption{\centering TOF=[112,120]ns\\ $\chi^2/NDF=5.29/4$}
    \end{subfigure}
    \hfill
    \begin{subfigure}[b]{0.45\textwidth}
        \includegraphics[width=1.0\textwidth]{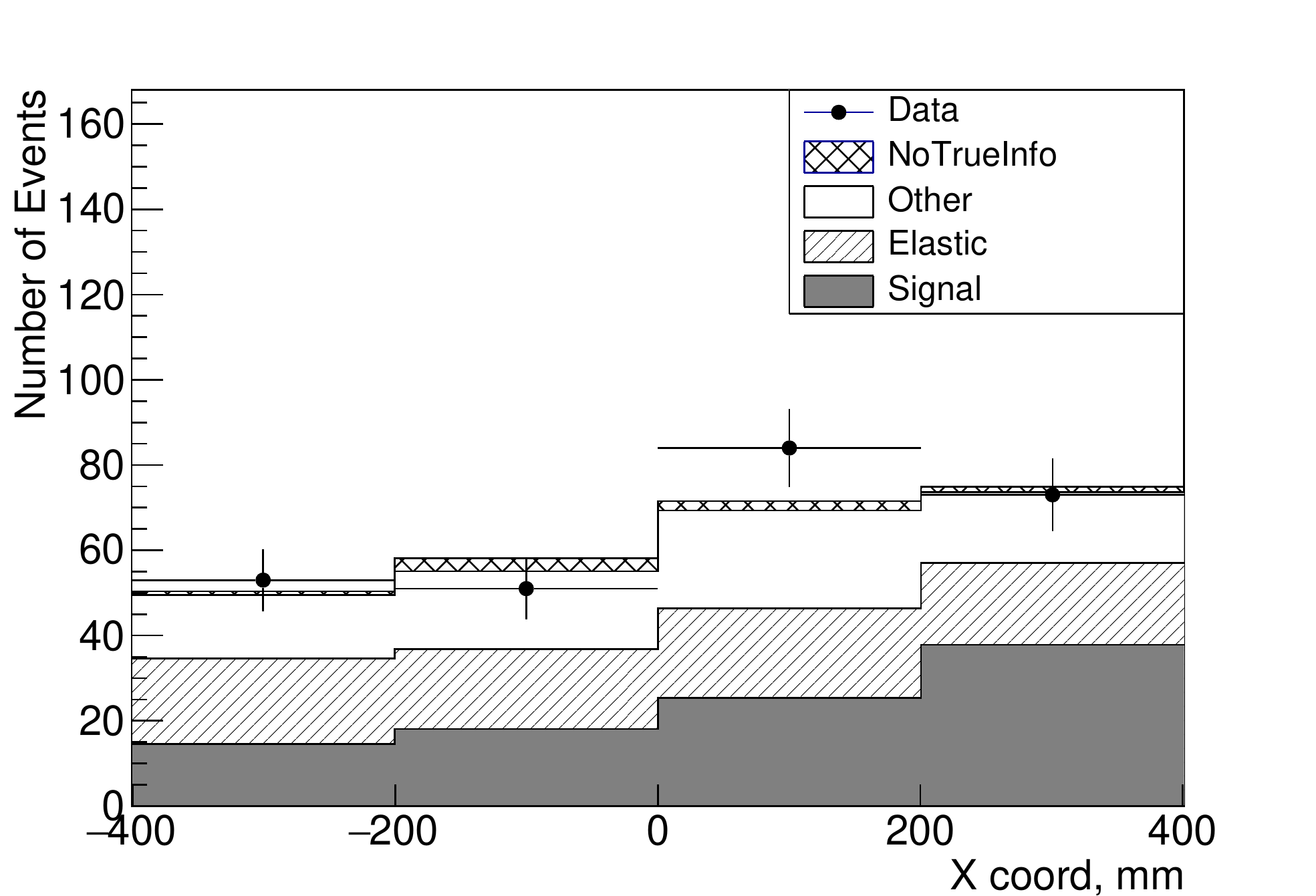}
        \caption{\centering TOF=[104,112]ns\\ $\chi^2/NDF=3.02/4$}
    \end{subfigure}
     \newline
      \begin{subfigure}[b]{0.5\textwidth}
      \centering
        \includegraphics[width=1\textwidth]{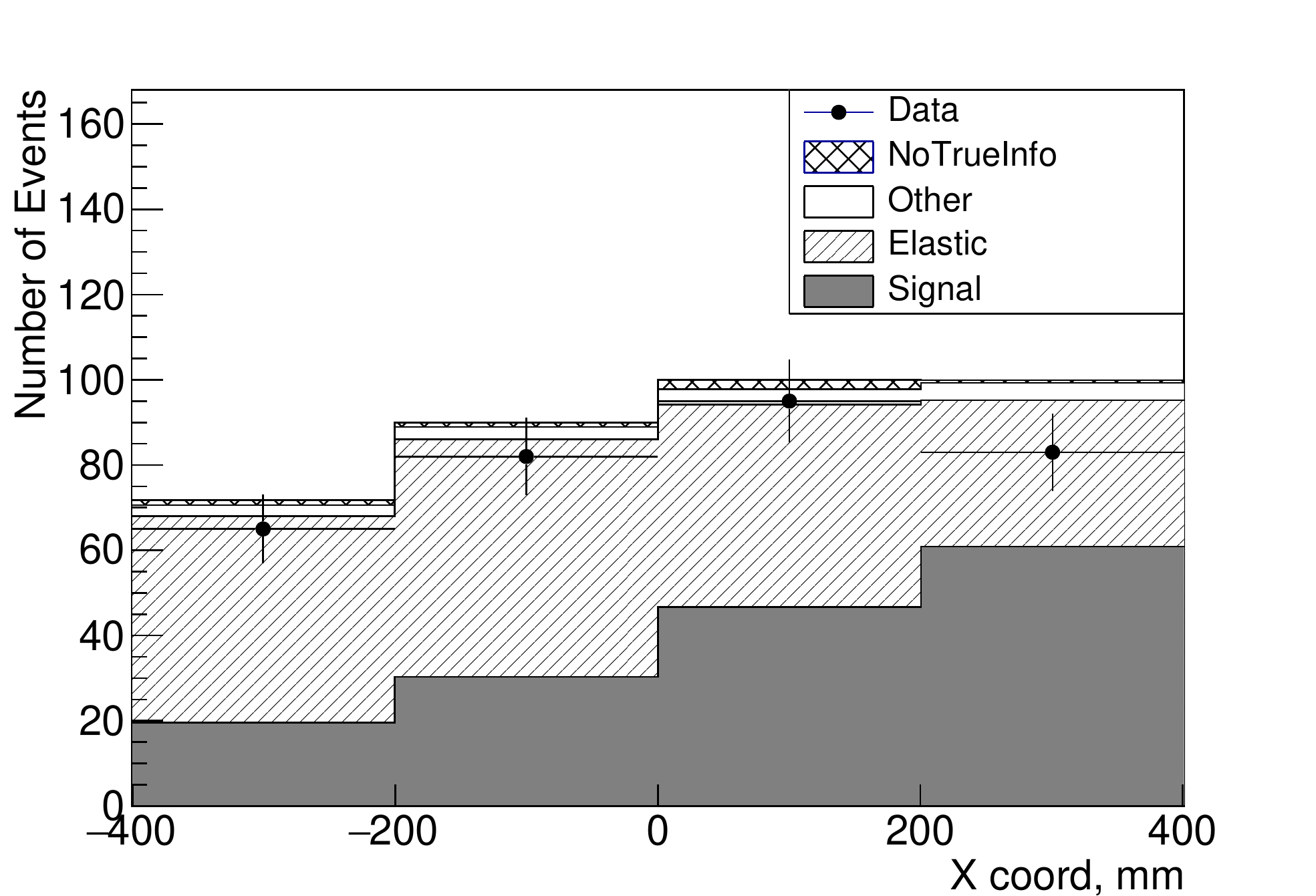}
        \caption{\centering TOF=[96,104]ns\\ $\chi^2/NDF=5.175/4$}
    \end{subfigure}

    \caption{Post-fit distribution of starting position of reconstructed tracks inside the side region for each TOF bin.}
    \label{fig:TOF_compare_side}
\end{figure*}

The fit demonstrates good agreement between the data and MC. The final value of $\chi^2$ is 42.12 with 36 degrees of freedom. Thus, the p-value is 0.223. The comparison between the data and posterior MC distributions for each TOF bin are presented in Fig.~\ref{fig:TOF_compare_sig} for the signal region and in Fig.~\ref{fig:TOF_compare_side} for the side region. 

\begin{figure}
\begin{center}
\includegraphics[width=0.46\textwidth]{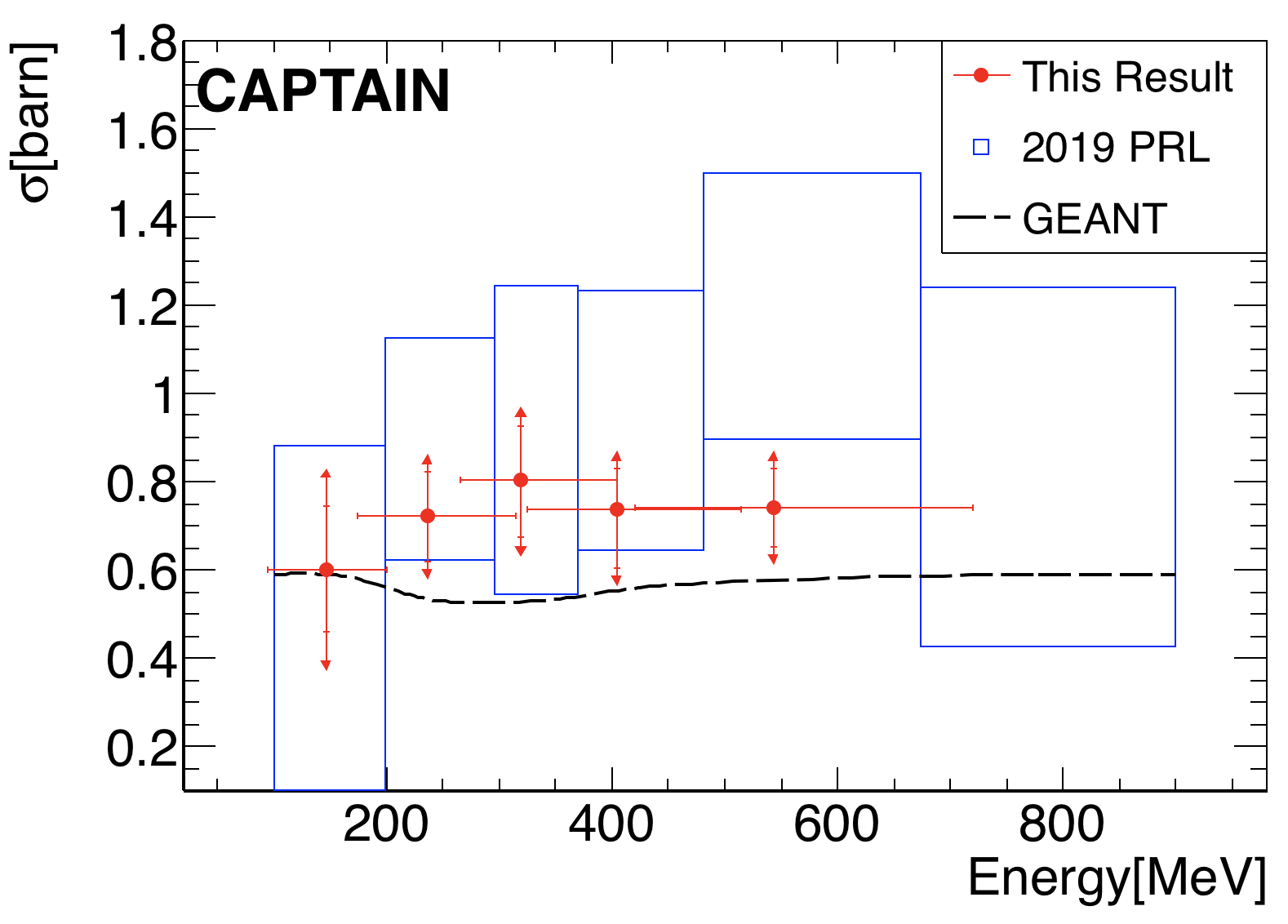}
\end{center}
\caption{The comparison between the initial neutron cross section measurement and the extended measurement. The blue squares represent the initial neutron cross section measurement with statistical and systematic errors added linearly. The red dots represent the extended cross section measurements for flux averaged energy in each of five TOF bins. The extended measurement is presented with statistical (lines) and systematic (arrows) errors. The black dashed line represents GEANT$\_$4.10.3 (QGSP$\_$BERT) base values of the cross sections used in the fit.}
\label{fig:CSMeasCompare}
\end{figure}

The final cross sections are given for flux averaged energies in the considered TOF bins: $\sigma(146~{\rm MeV})=0.60^{+0.14}_{-0.14}\pm0.08(\rm syst)$ b, $\sigma(236~{\rm MeV})=0.72^{+0.10}_{-0.10}\pm0.04(\rm syst)$ b, $\sigma(319~{\rm MeV})=0.80^{+0.13}_{-0.12}\pm0.040(\rm syst)$ b, $\sigma(404~\rm{MeV})=0.74^{+0.14}_{-0.09}\pm0.04(\rm syst)$ b, $\sigma(543~{\rm MeV})=0.74^{+0.09}_{-0.09}\pm0.04(\rm syst)$ b.  We find result of the extended measurement consistent with the result of the initial measurement of the CAPTAIN collaboration as shown in Fig.~\ref{fig:CSMeasCompare} and close to GEANT4 base values presented in Table~\ref{tab:base_cs}.

\section{Summary}
\label{sec:Summ}

In conclusion, we have presented the extended measurement of the beam depletion neutron cross section on argon between 95~MeV and 720~MeV. The measurement was obtained using the data from a 4.3-h exposure of the Mini-CAPTAIN detector to the WNR/LANSCE beam at Los Alamos National Lab in 2017. 

We carefully analyzed the uncertainty from systematic factors in the analysis and conclude that these effects are small compared to statistical uncertainties. The final fit shows a good agreement between data and MC with $\chi^2$/ndof=42.12/36. Moreover, we find the result consistent with the hypothesis of a small cross section change across the considered energy range. The $\chi^2$ for a given function with an average cross section of 0.721~b across all TOF bins gives value of 43.89. This value in combination with 36 degrees of freedom yields p-value of 0.172. 

The measurements presented here will provide more precise information to constrain the uncertainties of the current models of neutron transport. In turn, this will improve the neutrino energy reconstruction performance of liquid argon experiments attempting to resolve the CP violating phase and the neutrino mass hierarchy.

Research presented in this letter was supported by the Laboratory Directed Research and Development program of Los Alamos National Laboratory under project numbers 20120101DR and 20150577ER. This work benefited from the use of the Los Alamos Neutron Science Center, funded by the US Department of Energy under Contract No. DE-AC52-06NA25396 and we would like to
thank Nik Fotiadis, Hye Young Lee and Steve Wender for assistance with the 4FP15R beamline. We gratefully acknowledge the assistance of Mark Makela and the P25 neutron team. D.L.D. acknowledges his support as a Fannie and John Hertz Foundation Fellow, holding the Barbara Ann Canavan Fellowship. We further acknowledge the support of the US Department of Energy, Office of High Energy Physics and the University of Pennsylvania.

\bibliography{mybib}

\end{document}